\documentclass[conference]{IEEEtran}

\IEEEoverridecommandlockouts
\usepackage[switch]{lineno}
\usepackage{cite}
\usepackage[dvipsnames]{xcolor}
\usepackage{amsmath,amssymb,amsfonts}
\usepackage{algorithm}
\usepackage{algpseudocode}
\usepackage{graphicx}
\usepackage{textcomp}
\usepackage{subcaption}
\usepackage{bm}
\usepackage{stfloats}
\def\BibTeX{{\rm B\kern-.05em{\sc i\kern-.025em b}\kern-.08em
T\kern-.1667em\lower.7ex\hbox{E}\kern-.125emX}}
\usepackage{balance}

\begin{document}

\title{
Accelerating Mixed-Precision Out-of-Core Cholesky Factorization with Static Task Scheduling
}

\author{Jie Ren, Hatem Ltaief, Sameh Abdulah, and David E. Keyes
 \IEEEcompsocitemizethanks{\IEEEcompsocthanksitem The authors are with the Extreme Computing Research Center, Computer, Electrical, and Mathematical Sciences and Engineering (CEMSE) Division, King Abdullah University of Science and Technology (KAUST), Thuwal 23955-6900, Saudi Arabia. E-mail: {Jie.Ren, Hatem.Ltaief, Sameh.Abdulah, David.Keyes}@kaust.edu.sa.\protect\\}
}


\maketitle



\begin{abstract}
This paper explores the performance optimization of out-of-core (OOC) Cholesky factorization on shared-memory systems equipped with multiple GPUs. We employ fine-grained computational tasks to expose concurrency while creating opportunities to overlap data movement asynchronously with computations, especially when dealing with matrices that cannot fit on the GPU memory. We leverage the directed acyclic graph of the task-based Cholesky factorization and map it onto a static scheduler that promotes data reuse, while supporting strategies for reducing data movement with the CPU host when the GPU memory is exhausted. The CPU-GPU interconnect may become the main performance bottleneck as the gap between the GPU execution rate and the traditional PCIe bandwidth continues to widen. While the surface-to-volume effect of compute-bound kernels partially mitigates the overhead of data motion, deploying mixed-precision (MxP) computations exacerbates the throughput discrepancy. Using static task scheduling, we evaluate the performance capabilities of the new ultra-fast NVIDIA chip interconnect technology, codenamed NVLink-C2C, that constitutes the backbone of the NVIDIA Grace Hopper Superchip (GH200), against a new four-precision (FP64/FP32/FP16/FP8) \textit{left-looking} Cholesky factorization.
We report the performance results of a benchmarking campaign on various NVIDIA GPU generations and interconnects. We highlight 20\% performance superiority against cuSOLVER on a single GH200 with FP64, while hiding the cost of OOC task-based Cholesky factorization and we scale almost linearly on four GH200 superships. With MxP enabled, our statically scheduled four-precision tile-based Cholesky factorization scores a $3$X performance speedup against its FP64-only counterpart, delivering application-worthy FP64 accuracy when modeling a large-scale geospatial statistical application.
Our results represent a significant advance in handling large datasets for modern tightly coupled computing environments, especially for OOC applications requiring intensive data movement and processing.
\end{abstract}



\section{Introduction}
\label{sec:introduction}
The synergy between Central Processing Units (CPUs) and Graphics Processing Units (GPUs) in heterogeneous computing architectures has significantly propelled computational capabilities. A paramount challenge is the effective management of data movement, particularly in OOC scenarios for large-scale applications where the dataset size exceeds the GPU's onboard memory capacity. Traditional data transfer links, like Peripheral Component Interconnect Express (PCIe), have long been a bottleneck due to limited bandwidth and high latency, particularly in situations requiring frequent CPU-GPU data transfers. The widening gap between the GPU computational power and the PCIe bus bandwidth has further exacerbated this challenge due to loosely connected host-device hardware ecosystem. The AMD Accelerated Processing Unit (APU) was the first to combine a general-purpose x86 CPU with an integrated GPU on a single die, enabling both CPU and GPU to access the memory with the same address space~\cite{bouvier2014kabini}. Supported mainly by OpenCL, the programmability of the system allows for efficient and flexible utilization of heterogeneous computing resources. The NVLink technology substantially outperforms PCIe~\cite{elster2022nvidia}. For instance, the latest NVLink-C2C 4.0 technology can provide up to 900 GB/s of bandwidth, which overwhelms PCIe 4.0's 32 GB/s per lane. Therefore, NVLink provides a fresh opportunity to alleviate bottlenecks in OOC numerical algorithms. 

We demonstrate the ability of NVLink on the Grace Hopper GH200 Superchip to mitigate the overheads of data movement between the host and the device during the OOC Cholesky factorization of large dense matrices. By leveraging fine-grained task computations, we increase the degree of parallelism while enabling asynchronous data copying to improve performance on shared-memory systems equipped with multiple GPUs. The directed acyclic graph (DAG) of the tile-based Cholesky factorization that captures tasks (nodes) and their data dependencies (edges) can be traversed in two ways: (a) the right-looking variant where all tiles of the trailing submatrix are accessed and engaged in computations at the cost of activating expensive collective communications and (b) the left-looking variant where all tasks on tiles along the critical path are prioritized and launched before releasing the tiles once they reach their final state. The former is proactive and keeps busy all processing units of the system. The latter is lazy and may under-utilize the hardware resources in favor of prioritizing data reuse. Given the recursive formulation of the Cholesky factorization, we can map its DAG onto a deterministic static task scheduler that adheres to the left-looking technique, promotes critical data reuse, and reduces expensive host/device data movement. 

A typical static scheduler may assume all data is GPU resident when orchestrating tasks and does not provide data management in case the GPU memory runs out. Therefore, we augment the capabilities of our static scheduler for the left-looking Cholesky with support for OOC algorithms. Asynchronous CPU/GPU data movement techniques overlapped with useful computations and became critical for performance. Moreover, we deploy our static runtime system to perform mixed-precision (MxP) Cholesky factorization and support parametric geospatial data modeling applications~\cite{abdulah2018exageostat}. After determining the precision of the arithmetic for each tile, we launch a new four-precision (FP64/FP32/FP16/FP8) left-looking Cholesky factorization to take advantage of the application resiliency with respect to releasing trailing digits of the mantissa of low norm tiles, following the prescription of \cite{higham_mary_2022}. This requires our static scheduler to perform on-the-fly data type up/down-casting, further optimizing data movement between processing units by transmitting the minimum acceptable bytes per word.
In summary, our work can be distilled into the following contributions: (a) implement a static task scheduler that supports OOC computations, while leveraging data locality to minimize data transfer across multiple GPUs, (b) deploy data caching techniques to further reduce data movement and reuse previously accessed tiles left purposely in GPU memory, (c) adopt various floating-point precisions, namely FP64, FP32, FP16, and FP8, to take advantage of low-precision tensor core hardware optimizations, (d) evaluate the performance of the interconnect technologies, i.e., from PCIe to NVLink-C2C, using various GPU hardware generations, (e) demonstrate our efficient approach on spatial statistics applications to achieve significant acceleration while preserving the required accuracy levels, (f) achieve 20\% performance superiority against cuSOLVER on a single GH200 with FP64 and almost linearly scale on four GH200 superships, (g) score a 3X performance speedup when MxP enabled against its FP64-only counterpart, delivering application-worthy FP64 accuracy when modeling a large-scale geospatial statistical application.

The remainder of the paper is as follows. Section~\ref{sec:rw} presents related work and positions our research contributions. We recall the necessary background
in Section~\ref{sec:bg}. Section~\ref{sec:impl} provides implementation details on the static scheduler to support OOC algorithms, MxP, and multiple GPUs. We report performance results in Section~\ref{sec:results} and conclude in Section~\ref{sec:summary}.

\section{Related Work}
\label{sec:rw}



{\bf High-Bandwidth Interconnect Technologies.} The advent of high-bandwidth interconnects like NVLink has initiated new research efforts. GPU vendors have provisioned higher bandwidth and lower latency in interconnect technologies to mitigate data transfer overheads. 
Comparative studies highlight significant performance gains in data-intensive applications~\cite{li2019evaluating}. Additional research explores NVLink's role in out-of-device memory operations~\cite{appelhans2017leveraging, chien2019performance} and its integration into modern computational systems~\cite{chu2020nv}. A comprehensive case study of the Grace Hopper Superchip and NVLink interconnect was recently presented in \cite{fusco2024understanding}. The study highlights the significant impact of memory placement on performance using memory speed-focused micro-benchmarks on a GH200 quad node of the CSCS Alps supercomputer\footnote{https://www.cscs.ch/computers/alps}. The results show that the tightly integrated architecture of GH200, equipped with high-speed NVLink and C2C interconnects, provides a great ability to accelerate different workloads. This creates new scaling opportunities for memory-intensive applications.



{\bf Out-of-Core Matrix Algorithms on Heterogeneous Systems.}
The literature on enhancing the efficiency of matrix computations when the matrix size surpasses the onboard memory capacity of GPUs is rich. 
Studies have addressed this challenge by using asynchronous operations~\cite{rennich2016accelerating, ghysels2022high}. For instance, in \cite{rennich2016accelerating}, the authors partition the matrix into branches, i.e., subtrees terminating in leaves from the elimination tree. Each branch is sequentially loaded into the GPU memory, and all associated operations are executed on the GPU, eliminating the need for PCIe communication during computation. For larger matrices, the branches are streamed one after another through the GPU, enabling out-of-core (OOC) execution by accommodating matrices that are too large to fit entirely in GPU memory. Recently, fast interconnect technologies have advanced the execution of several applications~\cite{appelhans2017leveraging, chien2019performance, foley2017ultra}.

{\bf Mixed-Precision Computations.}
The use of MxP in scientific computing, particularly in linear algebra and numerical solvers, has received significant attention~\cite{Wilkinson-mp, Moler-mp, Dongarra-mp, haidar2018harnessing, amestoy_2021, higham_mary_2022, abdelfattah2021survey, abdelfattah2019fast}. The introduction of low-precision factorization, especially in the context of the Cholesky algorithm, represents a novel advancement in this field~\cite{abdelfattah2020investigating, higham2021exploiting, hogg2010fast}. Recently, MxP techniques have been utilized in linear algebra to enhance exascale scientific applications by relying on dynamic runtime systems. For instance, ~\cite{abdulah2024boosting} employs a three-level precision approach to demonstrate the first exascale climate emulator capable of producing spatial results at up to 3.5 km resolution for existing Earth models. Moreover, ~\cite{ltaief2024toward} demonstrates the use of four precisions to accelerate the performance of output accuracy-preserving MxP computations in Genome-Wide Association Studies (GWAS) involving $305K$ patients from the UK Biobank.


{\bf Geospatial Statistics Applications.}
Many studies have addressed the challenges of large-scale geospatial data modeling using numerical approximations, such as covariance tapering~\cite{furrer2006covariance,sang2012full}, likelihood approximations in both spatial~\cite{stein2004approximating} and spectral~\cite{fuentes2007approximate} domains.
See~\cite{sun2012geostatistics} for a comprehensive review.
More recent studies have explored the balance between computational speed and accuracy in MxP configurations~\cite{abdulah2019geostatistical, abdulah2021accelerating}, and the implications of these approaches for large-scale geospatial applications~\cite{abdulah2018exageostat,salvana2022parallel,abdulah2019geostatistical, abdulah2021accelerating, cao2022framework, cao2023reducing}.

{\bf Positioning this work.} In this paper, we augment our static scheduler for the Cholesky factorization with OOC features using data caching strategies by fetching data from device memory and host memory. We leverage the recursive formulation of the Cholesky factorization and deploy a left-looking variant to force a certain task execution order for maximizing data reuse, while controlling the application's memory footprint, thanks to the deterministic behavior of the static scheduler. This is something difficult to achieve with a dynamic runtime system using the right-looking variant due to its eager nature to expose parallelism at the cost of subsequently accessing several times the same tile. We bring the MxP capabilities to our static scheduler, combining four precisions (as low as FP8, depending upon the data) and still deliver FP64 application-worthy precision for geospatial statistics applications on multiple GPUs.

\section{Background}
\label{sec:bg}
This section presents an overview of the tile-based Cholesky factorization with the left-looking variant, powered by a static task scheduler.

\subsection{Left-Looking Tile Cholesky Factorization}
\begin{figure}[!ht]
\begin{subfigure}[b]{0.41\linewidth}
    \includegraphics[width=1.05\linewidth]{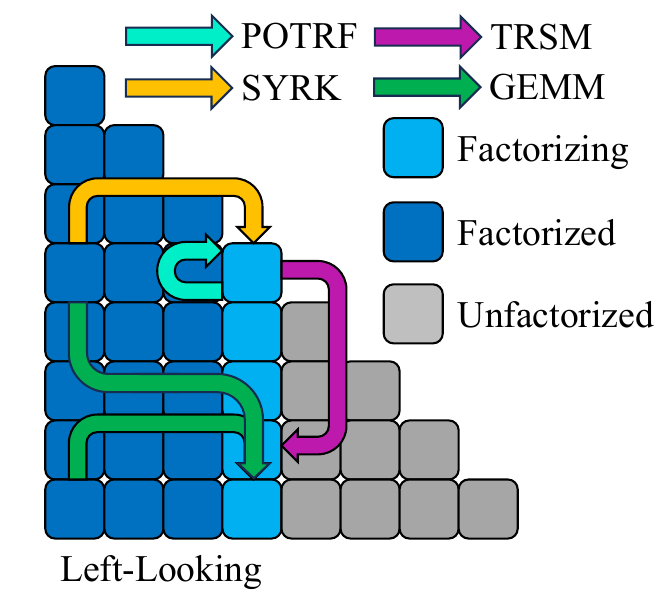}
    \caption{Four routines: POTRF, TRSM, SYRK, and GEMM.}
    \label{fig:method-left_looking}
\end{subfigure}
\hfill
\begin{subfigure}[b]{0.45\linewidth}
    \includegraphics[width=\linewidth]{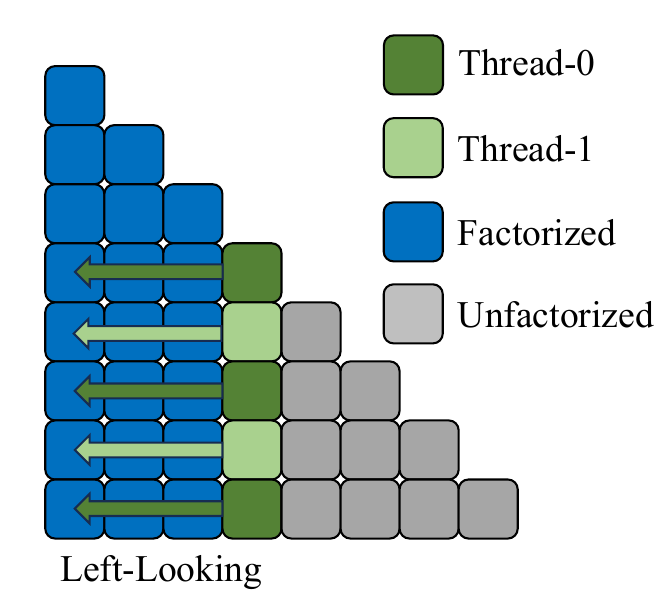}
    \caption{1D block-cyclic distribution and thread management.}
    \label{fig:method-static_scheduler}
\end{subfigure}
    \caption{Left-looking tile Cholesky with static scheduler.}
\end{figure}

A symmetric positive-definite (SPD) matrix $A$ can be factorized as $A = LL^T$, where $L$ is the lower triangular Cholesky factor. 
To leverage fine-grained computations and expose concurrency, we divide the matrix $A$ into an array of tiles $A_{i, j}$, with $i$ and $j$ representing the tile's row and column indices. The left-looking variant traverses the tiles of $A$ column by column, as illustrated in Figure~\ref{fig:method-left_looking}. A diagonal tile $A_{i, i}$ is first updated by considering all tiles to its left using the formula $A_{i, i} = -\sum_{k=0}^{i-1} A_{i, k}A_{i, k}^T + A_{i, i}$ (SYRK) and then factorized into a triangular matrix $L_{i, i}$ (POTRF), where $A_{i, i} = L_{i, i}L_{i, i}^T$. An off-diagonal tile $A_{i, j}$ is updated in a similar manner with the formula $A_{i, j} = -\sum_{k=0}^{i-1} A_{i, k}A_{j, k}^T + A_{i, j}$ (GEMM) and then solved to obtain $L_{i, j}$, where $L_{i, j}L_{i, i}^T = A_{i, j}$ (TRSM).

{\small
\begin{algorithm}[!ht]
\caption{Left-looking tile Cholesky factorization with static scheduler, the loops in \textcolor{blue}{blue} are iterations with tasks assigned to threads in a 1D cyclic manner.}
\begin{algorithmic}[1]
\Require $\boldsymbol{A}$, where $\boldsymbol{A}$ is a matrix of size $n \times n$, partitioned into $Nt \times Nt$ tiles.
\State Initialize dependency table $Ready = \{\}$.
\For{$\textcolor{blue}{k = 0:Nt}$} \Comment{Parallel tasks}
    \For{$\textcolor{blue}{m = k:Nt}$} \Comment{Parallel tasks}
        \If{$m == k$} \Comment{Diagonal}
            \For{$n = 0:k$} \Comment{Update, SYRK}
                \State Wait until $Ready[k, n]$ is $True$
                \State $\boldsymbol{A}_{k, k} = -\boldsymbol{A}_{k, n}\boldsymbol{A}_{k, n}' + \boldsymbol{A}_{k, k}$
            \EndFor
                \State $\boldsymbol{A}_{k, k}$ = POTRF($\boldsymbol{A}_{k, k}$) \Comment{Factorize, POTRF}
                \State Set $Ready[k, n] = True$
        \Else \Comment{Off-diagonal}
            \For{$n = 0:k$} \Comment{Update, GEMM}
                \State Wait until $Ready[m, n]$ is $True$
                \State Wait until $Ready[k, n]$ is $True$
                \State $\boldsymbol{A}_{m, k} = -\boldsymbol{A}_{m, n}\boldsymbol{A}_{k, n}' + \boldsymbol{A}_{m, k}$
            \EndFor
            \State Wait until $Ready[k, k]$ is $True$
            \State $\boldsymbol{A}_{m, k} = \boldsymbol{A}_{m, k} \backslash \boldsymbol{A}_{k, k}'$ \Comment{Factorize, TRSM}
            \State Set $Ready[m, k] = True$
        \EndIf
    \EndFor
\EndFor
\end{algorithmic}
\label{alg:chol}
\end{algorithm}
}

\subsection{Static Scheduler}

The static scheduler approach involves assigning tasks to each thread in a one-dimensional (1D) block-cyclic fashion~\cite{onesided-static,twosided-static,potrfgpu}. This method of task distribution is visually represented in Figure~\ref{fig:method-static_scheduler} and further detailed in Algorithm~\ref{alg:chol}.

The entire matrix factorization process is segmented into a series of tasks, with each task focusing on the update and factorization of a specific tile. A strength of this approach is that each thread is aware of its assigned tiles from the outset. We rely on busy waits on a progress table to avoid violations of data dependencies. This deterministic behavior allows for a streamlined execution of the factorization process.

Indeed, the predetermined nature of task assignments under this static scheduling model proves particularly advantageous in managing data movement, a critical aspect in GPU implementations. In such environments, optimizing data transfer between the CPU and GPU, as well as within the GPU memory, is crucial for performance. By knowing in advance which tiles each thread will work on, unnecessary data movements can be eliminated, thereby reducing overheads and enhancing the overall efficiency of the factorization process. This method promotes faster computation and better uses the computational resources available on GPUs.

\subsection{Out-of-Core Algorithms}
\label{subsec:ooc}

\begin{figure}[!ht]
\centering
\includegraphics[width=0.45\textwidth]{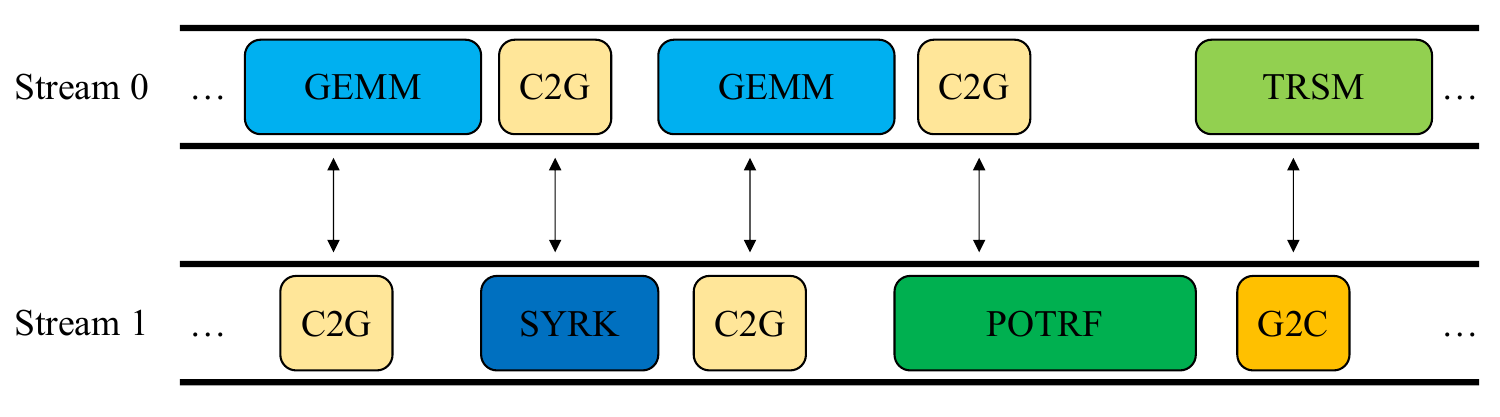}
\caption{Multiple streams are executed on each device to overlap data motion with computations (\textit{async} implementation).}
\label{fig:method-overlapping}
\end{figure}

NVIDIA introduced the concept of streams to enable more granular control over GPU operations, allowing for parallel execution and efficient management of tasks. A stream functions as a queue for tasks, allowing users to enqueue various operations, such as launching CUDA kernels or initiating memory transfers. These tasks are executed in the order received, adhering to a first-in-first-out (FIFO) principle. Users have the flexibility to allocate one or more streams on each GPU. This multi-stream allocation is particularly useful in managing and optimizing the execution of different tasks on the GPU. A notable advantage of using streams is the ability to overlap memory transfers between the CPU and GPU with ongoing computations on either the GPU or CPU.

The potential for reducing memory copying overhead in both directions through this method is significant, as illustrated in Figure~\ref{fig:method-overlapping}. To achieve this efficiency, however, a carefully designed pipeline is required. Such a pipeline must schedule tasks and manage memory to hide the latency associated with data transfers, thereby optimizing the overall performance of the application. This strategy is a key component of the OOC feature that leverages the full power of modern GPU architectures for high-performance computing tasks.

\subsection{Geospatial Data Modeling Application}
Gaussian processes (GPs) are widely used in geospatial analysis to model spatial data and predict unknown values in uncertain or incomplete information areas. The process involves using the Maximum Likelihood Estimation (MLE) technique to estimate statistical parameters. These parameters characterize the underlying spatial field and rely on a predefined covariance function. Assume there are $n$ spatial locations, ${\bf s}_1, \ldots, {\bf s}_n \in \mathbb{R}^d$ and that $\mathbf{y} = \left(y_1, \ldots, y_n\right)^\top$ represents an observation vector associated with different locations. In a Gaussian process, the data $\mathbf{y}$ are modeled as $\mathbf{y} \sim \mathcal{N}\left(\bm \mu, \bm \Sigma_{\boldsymbol\theta}\right)$, where $\bm \Sigma_{\boldsymbol\theta}$ is a covariance matrix with $(i, j)$ entry determined by a given covariance function $C_{\boldsymbol\theta}\left({\bf s}_i, {\bf s}_j\right)$ relying on a vector of covariance parameters,  ${\boldsymbol\theta}$. The log-likelihood estimation function can be represented as:
\begin{equation}
	\label{eq:likeli}
 \ell({\boldsymbol\theta};\boldsymbol{y})=-\frac{n}{2}\log(2\pi) - \frac{1}{2}\log |{{\boldsymbol \Sigma}_{\boldsymbol\theta}}|-\frac{1}{2}{\boldsymbol{y}}^\top {\boldsymbol \Sigma}_{\boldsymbol\theta}^{-1}{\boldsymbol{y}}.
\end{equation}

\noindent
The primary challenge in MLE, particularly with large datasets, is computing the $n^2$ elements of $\boldsymbol{\Sigma}_{\boldsymbol{\theta}}$ and its factorization, of cubic complexity in $n$. The covariance matrix $\bm \Sigma_{\boldsymbol\theta}$ is constructed based on a covariance function with spatial or temporal arguments. In this work, we employ the Mat\'ern covariance function~\cite{gneiting2010matern} with the form:  
\begin{equation}\label{eqn:matern}
C(\|\mathbf{h}\|; \boldsymbol{\theta}) = \frac{\sigma^2}{2^{\nu - 1} \Gamma \left( \nu \right)} \left(\frac{\|\mathbf{h}\|}{a}\right)^{\nu} \mathcal{K}_{\nu}\left(\frac{\|\mathbf{h}\|}{a}\right),
\end{equation}
where $\mathcal{K}_{\nu}(\cdot)$ represents the modified Bessel function of the second kind with order $\nu$, $\Gamma(\cdot)$ denotes the gamma function, $\mathbf{h}$ represents the Euclidean distance between two locations $s_i$ and $s_j$. $\boldsymbol{\theta}$ contains the marginal variance ($\sigma^2>0$), smoothness ($\nu>0$), and spatial range ($a>0$). 
Evaluating the log of the determinant, $\log |{{\boldsymbol \Sigma}({\boldsymbol\theta})}| $, and applying ${\boldsymbol \Sigma}({\boldsymbol\theta})^{-1}$ require computing the Cholesky factorization for ${\boldsymbol \Sigma}({\boldsymbol\theta})$. 

We employ the proposed MxP method to perform the factorization with different accuracy levels and assess the accuracy of the factorization in the obtained log-likelihood value. We use the Kullback-Leibler (KL) divergence metric to evaluate the loss of accuracy that may affect the performance of estimating the Gaussian 
log-likelihood~\cite{pan2024gpu}. The KL divergence is a statistical tool employed to quantify the discrepancy in accuracy between two probability distributions. Considering two statistical models, $A$ and $B$, KL divergence quantifies the information loss incurred when model $B$ is used instead of the actual model $A$. We apply the KL divergence criterion to measure the loss of information in estimated Gaussian log-likelihood when using MxP Cholesky factorization for decomposing $\Sigma$ compared to the full double-precision approach. The computation of KL divergence in this scenario is:
\begin{equation}
    D_{\text{KL}}({\mathcal {N}}_{0}\parallel {\mathcal {N}}_{a})= \ell_0({\boldsymbol\theta};\boldsymbol{0}) - \ell_a({\boldsymbol\theta};\boldsymbol{0}),
    \label{eq:kl-gaussian}
\end{equation}
\noindent
where $\ell_0({\boldsymbol\theta};\boldsymbol{0})$ is the exact log-likelihood at $\mathbf{y} = \boldsymbol{0}$ while $\ell_a({\boldsymbol\theta};\boldsymbol{0})$ is the Vecchia-approximated log-likelihood at $\mathbf{y} = \boldsymbol{0}$.




\section{Implementation Details}
\label{sec:impl}
This section details our proposed OOC implementations of the left-looking MxP Cholesky factorization.

\subsection{Optimizing Data Transfer and Memory Usage in CPU-GPU interactions: \textit{sync} vs \textit{async} implementations}
\label{subsec:pinned}
As shown in Figure~\ref{fig:method-overlapping}, multiple streams per device can help in optimizing the CPU-GPU data transfer. Our baseline implementation has two main versions: \textit{sync} and \textit{async}. Compared to the \textit{sync} implementation, the \textit{async} implementation uses multiple streams per device to effectively manage the CPU-GPU data transfer. Each stream is tasked with independently updating and factorizing the tiles assigned to it. This approach allows for simultaneous operations, where one stream can be engaged in computational tasks while another stream concurrently performs an asynchronous memory copy operation. Such a strategy is instrumental in concealing the data movement latency, which is particularly beneficial when using high-speed CPU-GPU bus interconnect networks.

An additional step to enhance the performance of the \textit{async} implementation is pinned memory instead of pageable memory. Pinned memory, also known as non-pageable memory, is a type of memory that remains fixed in the physical RAM and is not swapped to the disk by the operating system. Pinned memory reduces the latency associated with memory transfers between the CPU and GPU, thereby improving the overall data handling efficiency. The advantages of using pinned memory are even more significant on the Grace Hopper server. 

\subsection{Efficient Data Management in Cholesky Decomposition and GEMM Operations}
\begin{figure*}[!ht]
\centering
\begin{subfigure}[b]{0.31\linewidth}
\centering
    \includegraphics[width=0.8\linewidth]{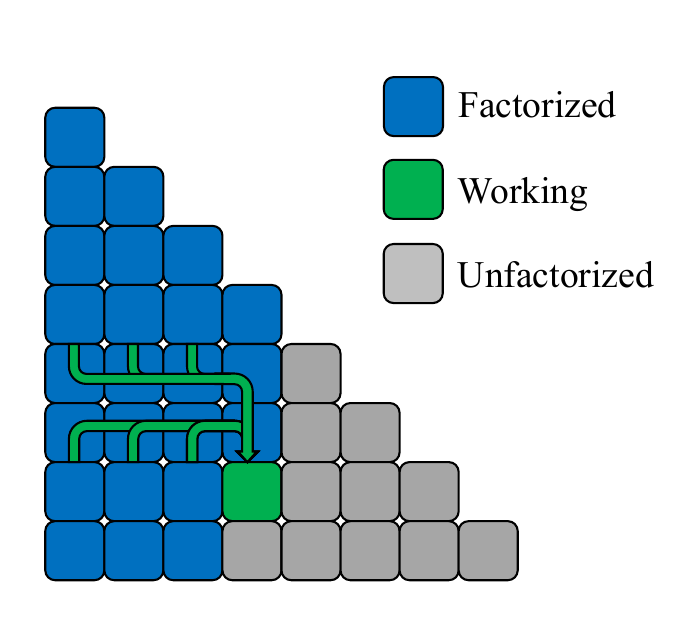}
    \caption{V1: For each stream, the accumulator ($C$ in $C = -AB^T + C$) remains in memory.}
    \label{fig:method-V1}
\end{subfigure}
\hfill
\begin{subfigure}[b]{0.31\linewidth}
\centering
    \includegraphics[width=0.8\linewidth]{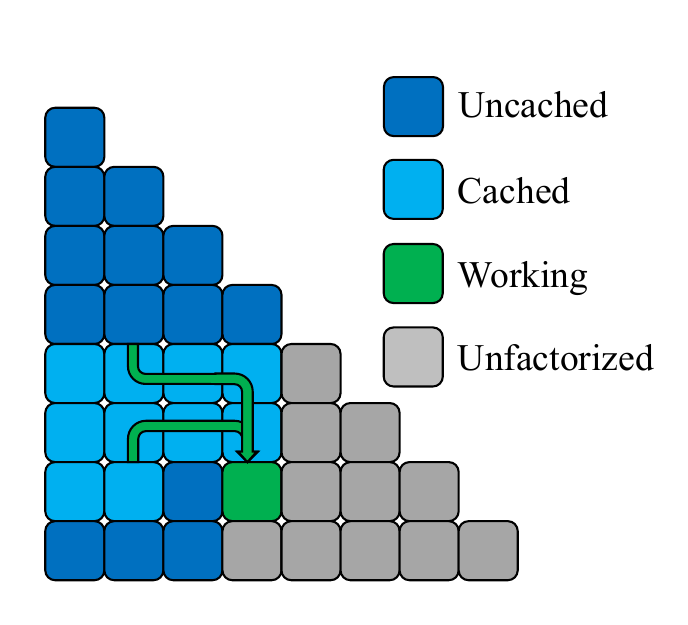}
    \caption{V2: repurposing the space of unused tiles when GPU memory is limited.}
    \label{fig:method-V2}
\end{subfigure}
\hfill
\begin{subfigure}[b]{0.31\linewidth}
\centering
    \includegraphics[width=0.8\linewidth]{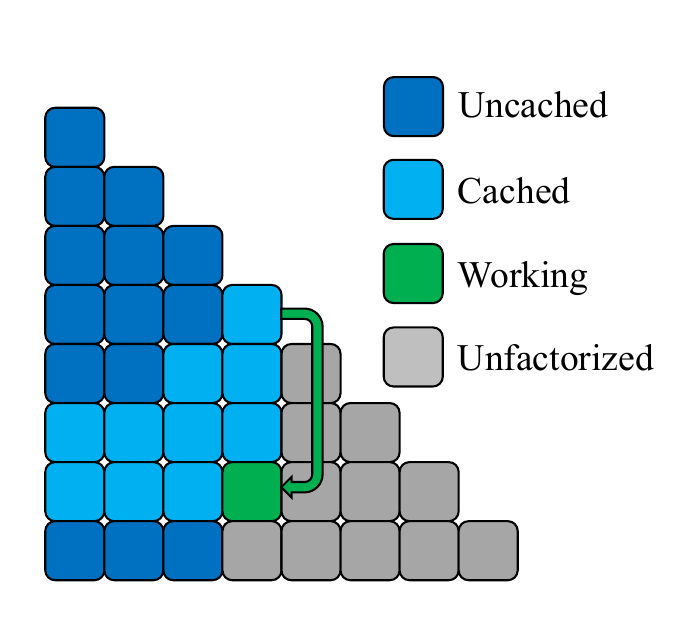}
    \caption{V3: the diagonal tile is kept until all tiles of a column block complete TRSM tasks.}
    \label{fig:method-V3}
\end{subfigure}
\caption{Strategies for further reducing data movement of the \textit{async} implementation with three different versions.}
\label{fig:three_versions}
\end{figure*}

To further optimize the \textit{async} implementation, we introduce three refined versions of the left-looking Cholesky decomposition algorithm, as shown in Figure~\ref{fig:three_versions}. We rely on static scheduling in these versions to perform the Cholesky decomposition more efficiently. For a given stream, while updating a tile, the accumulator (denoted as $C$ in the equation $C = -AB^T + C$) remains constant. This is illustrated in Figure~\ref{fig:method-V1} and Algorithm~\ref{alg:memory-optimization}, where the \textcolor{Green}{green tile} represents the accumulator. We avoid loading the accumulator multiple times, significantly reducing data transfer volume to the GPU. We call this version \textit{V1} for the rest of the paper.

Furthermore, we extended the \textit{V1} implementation with an additional progress table to reduce data volume when copying GEMM operands ($A$ and $B$ in $C = -AB^T + C$). This table contains pointers to matrix tiles stored in GPU memory. Before executing GEMM, we consult the table as Algorithm~\ref{alg:load_tile} shows: if the required matrix tile is present, we reuse it; if not, it is copied from the CPU and added to the table.  Once the GEMM operation terminates, we retain the operands' space for subsequent operations instead of freeing them in the table.  When GPU memory becomes limited, we repurpose the space of least or non-utilized tiles. This version, referred to as \textit{V2} and shown in Figure~\ref{fig:method-V2}, illustrates the reuse of operand $B$ to further reduce data movement.

{\small
\begin{algorithm}[!ht]
\caption{GPU implementation for Algorithm~\ref{alg:chol},  the \textcolor{blue}{blue} loops are assigned to threads in a 1D cyclic manner, \textcolor{Green}{green tiles} are reused accumulators, \textcolor{orange}{orange tiles} are reused diagonal tiles. Algorithm~\ref{alg:load_tile} is used to cache and reuse other tiles.}
\begin{algorithmic}[1]
\Require $\boldsymbol{A}$, where $\boldsymbol{A}$ is a matrix of size $n \times n$, partitioned into $Nt \times Nt$ tiles.
\State Initialize dependency table $Ready = \{\}$ and cache table $Cache = \{\}$.
\For{$\textcolor{blue}{k = 0:Nt}$} \Comment{Parallel tasks}
    \For{$\textcolor{blue}{m = k:Nt}$} \Comment{Parallel tasks}
        \If{$m == k$} \Comment{Diagonal}
            \State $\textcolor{Green}{\boldsymbol{A}_{k, k}} = load\_tile(Cache, k, k)$
            \For{$n = 0:k$} \Comment{Update, SYRK}
                \State Wait until $Ready[k, n]$ is $True$
                \State $\boldsymbol{A}_{k, n} = load\_tile(Cache, k, n)$
                \State $\textcolor{Green}{\boldsymbol{A}_{k, k}} = -\boldsymbol{A}_{k, n}\boldsymbol{A}_{k, n}' + \textcolor{Green}{\boldsymbol{A}_{k, k}}$
            \EndFor
                \State $\textcolor{Green}{\boldsymbol{A}_{k, k}}$ = POTRF($\textcolor{Green}{\boldsymbol{A}_{k, k}}$) \Comment{Factorize, POTRF}
                \State $memcpy\_GPU\_to\_CPU(k, k, \textcolor{Green}{\boldsymbol{A}_{k, k}})$
                \State Set $Ready[k, k] = True$
        \Else \Comment{Off-diagonal}
            \State $\textcolor{Green}{\boldsymbol{A}_{m, k}} = load\_tile(Cache, m, k)$
            \For{$n = 0:k$} \Comment{Update, GEMM}
                \State Wait until $Ready[m, n]$ is $True$
                \State $\boldsymbol{A}_{m, n} = load\_tile(Cache, m, n)$
                \State Wait until $Ready[k, n]$ is $True$
                \State $\boldsymbol{A}_{k, n} = load\_tile(Cache, k, n)$
                \State $\textcolor{Green}{\boldsymbol{A}_{m, k}} = -\boldsymbol{A}_{m, n}\boldsymbol{A}_{k, n}' + \textcolor{Green}{\boldsymbol{A}_{m, k}}$
            \EndFor
            \State Wait until $Ready[k, k]$ is $True$
            \State $\textcolor{orange}{\boldsymbol{A}_{k, k}} = load\_tile(Cache, k, k)$
            \State $\textcolor{Green}{\boldsymbol{A}_{m, k}} = \textcolor{orange}{\boldsymbol{A}_{k, k}}' \backslash \textcolor{Green}{\boldsymbol{A}_{m, k}}$ \Comment{Factorize, TRSM}
            \State $memcpy\_GPU\_to\_CPU(m, k, \textcolor{Green}{\boldsymbol{A}_{m, k}})$
            \State Set $Ready[m, k] = True$
        \EndIf
    \EndFor
\EndFor
\end{algorithmic}
\label{alg:memory-optimization}
\end{algorithm}
}

{\small
\begin{algorithm}[]
\caption{$load\_tile$: Tile Loading with Cache Table.}
\begin{algorithmic}[1]
\Require $Cache$, $m$, and $n$, where $Cache$ is the data table, $m$ is the row index of the tile, and $n$ is the column index of the tile.
\Ensure $p$, a pointer to the GPU address of tile $A_{m, n}$.
\If{$exist(Cache, m, n)$} \Comment{Check existence}
    \State $p = get(Cache, m, n)$ \Comment{Get the cached tile's pointer}
\Else
    \If{$oom(Cache, m, n)$} \Comment{Out of Memory}
        \State $remove\_steal(Cache)$
        \State $p = load\_tile(Cache, m, n)$
    \Else \Comment{Allocate new}
        \State $p = malloc\_GPU()$
        \State $memcpy\_CPU\_to\_GPU(m, n, p)$ \Comment{Cache tile}
        \State $insert(Cache, m, n, p)$
    \EndIf
\EndIf
\end{algorithmic}
\label{alg:load_tile}
\end{algorithm}
}
Additionally, since each tile within a column block requires the corresponding diagonal tile for TRSM, we have modified the data table, as shown in Figure~\ref{fig:method-V3}. This ensures the diagonal tile does not get flushed out when GPU memory is scarce until all the tiles in a column block have completed their TRSM tasks. We refer to this version as \textit{V3}.

\subsection{Mixed Precisions for Efficiency and Memory Optimizations}
\begin{figure*}[!ht]
\begin{subfigure}[b]{0.18\linewidth}
    \includegraphics[width=\linewidth]{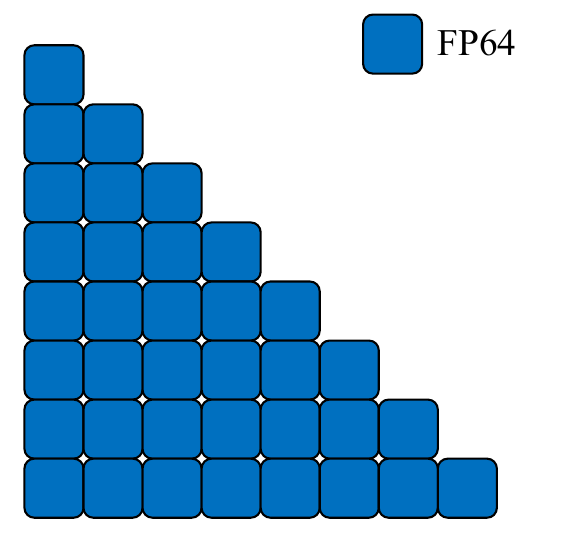}
    \caption{One precision.}
    \label{fig:method-FP64}
\end{subfigure}
\hfill
\begin{subfigure}[b]{0.18\linewidth}
    \includegraphics[width=\linewidth]{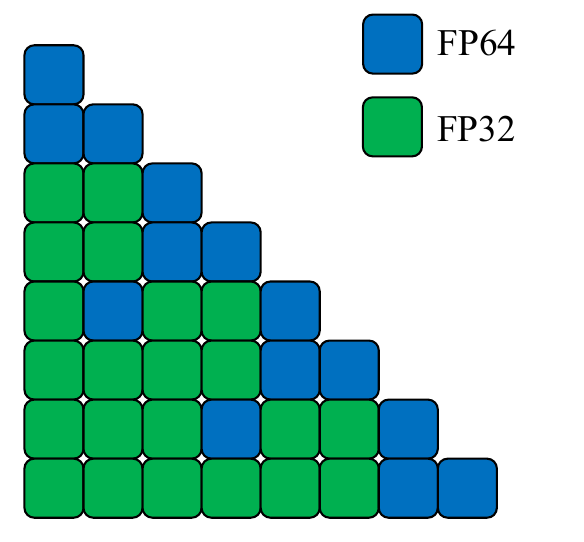}
    \caption{Two precisions.}
    \label{fig:method-FP64_FP32}
\end{subfigure}
\hfill
\begin{subfigure}[b]{0.18\linewidth}
    \includegraphics[width=\linewidth]{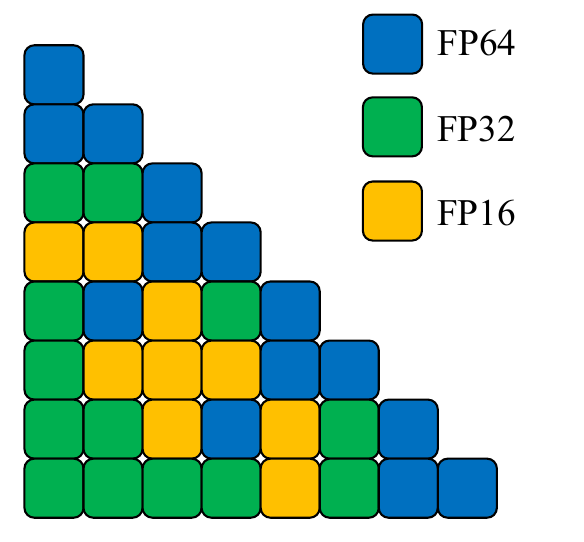}
    \caption{Three precisions.}
    \label{fig:method-FP64_FP32_FP16}
\end{subfigure}
\hfill
\begin{subfigure}[b]{0.18\linewidth}
    \includegraphics[width=\linewidth]{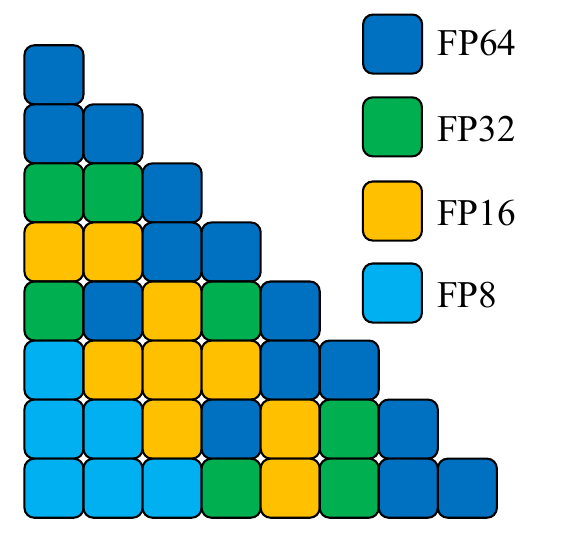}
    \caption{Four precisions.}
    \label{fig:method-FP64_FP32_FP16_FP8}
\end{subfigure}
\caption{MxP management on a 8x8 tile matrix.}
\label{fig:method-mixed_precision}
\end{figure*}
To enhance computational speed while minimizing memory usage, we explore low-precision arithmetic for matrix tiles. Our approach aligns with the threshold criteria outlined in \cite{higham_mary_2022} for determining the adaptive precision of each tile. Initially, we calculate the Frobenius norm of the entire matrix and then for each tile. We assess whether the inequality $\frac{n||A_{i,j}||_F}{||A||_F} < \frac{\epsilon_{high}}{\epsilon_{low}}$ holds, where $\epsilon_{low}$ represents the machine epsilon for the desired precision level, $n$ is the tile number of each column block and $i, j$ represent the tile index. If this inequality is satisfied, we consider a lower precision; otherwise, we opt for a higher precision. This strategy enables us to downgrade the precision of certain tiles to as low as FP8. Figure~\ref{fig:method-mixed_precision} illustrates the selection of precision levels for matrix tiles under various precision combinations. 

We introduce four precisions with our optimized static scheduler for geospatial modeling, unlike previous work \cite{abdulah2019geostatistical,abdulah2021accelerating,cao2023reducing} that used the PaRSEC dynamic runtime system. 

\subsection{Implementation on Multiple GPUs}
\begin{figure}[!ht]
\begin{subfigure}[b]{0.4\linewidth}
    \includegraphics[width=1.08\linewidth]{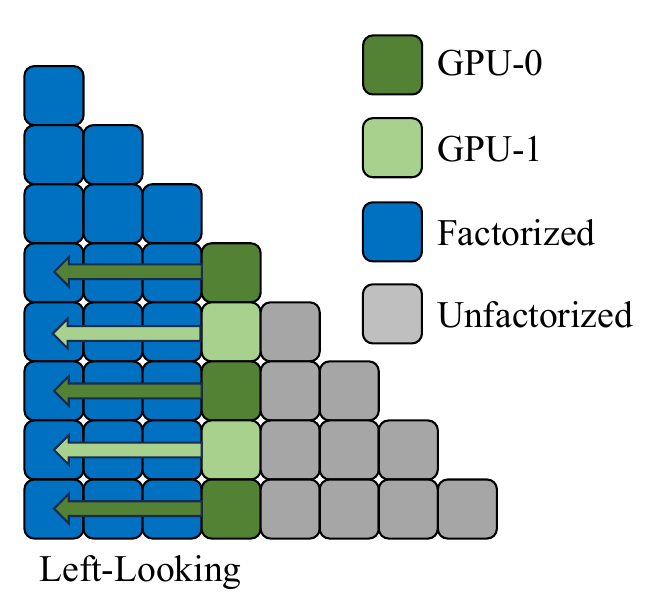}
    \caption{Illustration of 1D block-cyclic task distribution across devices for balanced workload.}
    \label{fig:method-dist_work}
\end{subfigure}
\hfill
\begin{subfigure}[b]{0.45\linewidth}
    \includegraphics[width=\linewidth]{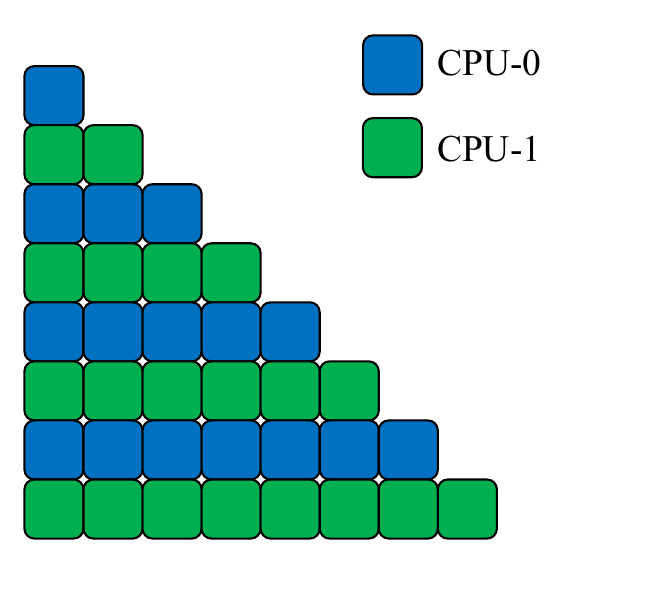}
    \caption{Depiction of 1D block-cyclic memory allocation on the Grace Hopper server for load balance.}
    \label{fig:method-dist_memory}
\end{subfigure}
\caption{Distributed algorithm and memory management.}
\end{figure}

The distributed implementation of our algorithm is depicted in Figure~\ref{fig:method-dist_work}. Assuming using two GPUs, we ensure that the tiles assigned to each GPU are distributed in a 1D block-cyclic fashion to trigger a balanced workload distribution across the devices.

Our multi-GPU implementation considers the NUMA architecture of the new Grace Hopper GH200 Superchip server.
This hardware configuration exhibits varying data access speeds when interfacing with different CPU memories. Notably, the data transfer bandwidth from a CPU to its directly attached GPU can reach up to 900 GB/s, achievable only with pinned memory, as described in Section~\ref{subsec:pinned}. In contrast, this bandwidth significantly drops to a maximum of 100 GB/s for GPUs that are not directly connected to a CPU. Data needs to be strategically allocated across different CPU memories to mitigate these speed variances and ensure load balancing. This allocation must also adhere to a 1D block-cyclic pattern to match the static scheduler task execution order, as depicted in Figures~\ref{fig:method-static_scheduler} and~\ref{fig:method-dist_memory}.

\section{Experimental Results}
\label{sec:results}

We conduct a set of experiments with five objectives: (1) to benchmark the proposed OOC Cholesky implementations on single and multiple recent NVIDIA GPU devices, (2) to evaluate data movement between CPU and GPU across different GPU architectures, (3) to evaluate the accuracy of MxP approach for geospatial data modeling, (4) to assess the performance of applying FP64, FP32, FP16, and FP8 MxP computations in the context of geospatial statistics applications, and (5) to highlight the advantages of the fast NVLink-C2C interconnect within the NVIDIA GH200 superchip compared to PCI interconnect. We use several platforms: 4xA100-PCIe 80GB (Intel Xeon CPU with 1TB of memory), 4xH100-PCIe 80GB (Intel Xeon CPU with 1TB of memory), and a pre-production 4xGH200 Grace Hopper Superchip 80GB system (ARM CPU with 480GB of memory). The A100-PCIe and H100-PCIe servers are connected via fourth—and fifth-generation PCIe buses with four lanes, while the GH200-NVL-C2C servers utilize fourth-generation NVLink buses for CPU-GPU communication. 

\subsection{Single GPU Performance}
In our first set of experiments, we evaluate the performance of our proposed left-looking Cholesky factorization implementation using FP64 arithmetic on a single GPU.
\subsubsection{Single GPU Flop/s Comparison}
\begin{figure*}[!ht]
\begin{subfigure}[b]{0.24\linewidth}
    \includegraphics[width=\linewidth]{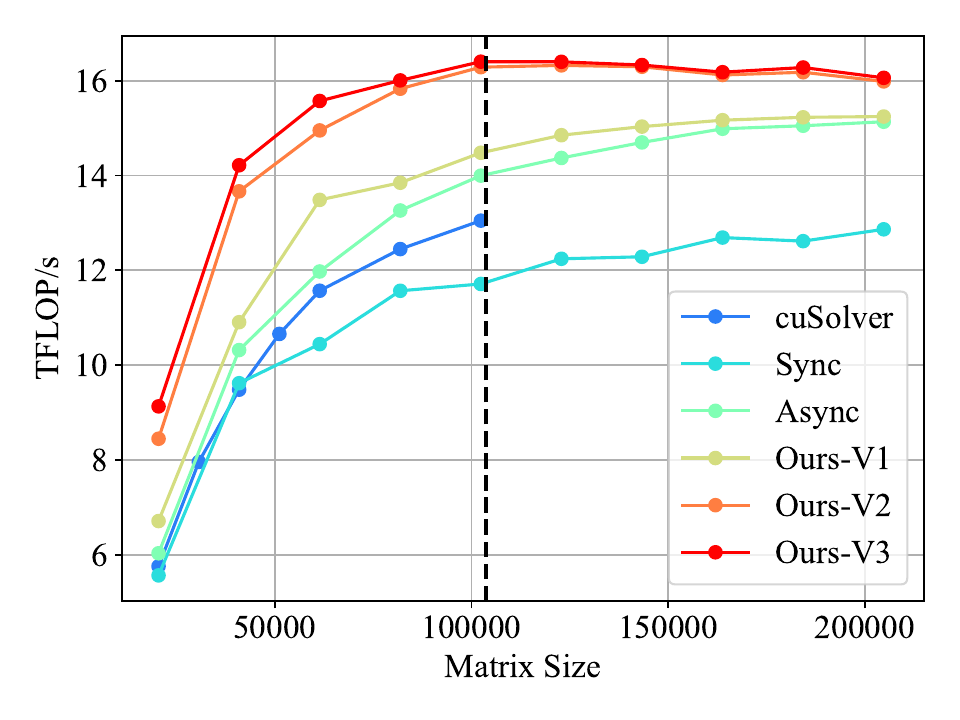}
    \caption{A100 with PCIe Gen4.}
\end{subfigure}
\hfill
\begin{subfigure}[b]{0.24\linewidth}
    \includegraphics[width=\linewidth]{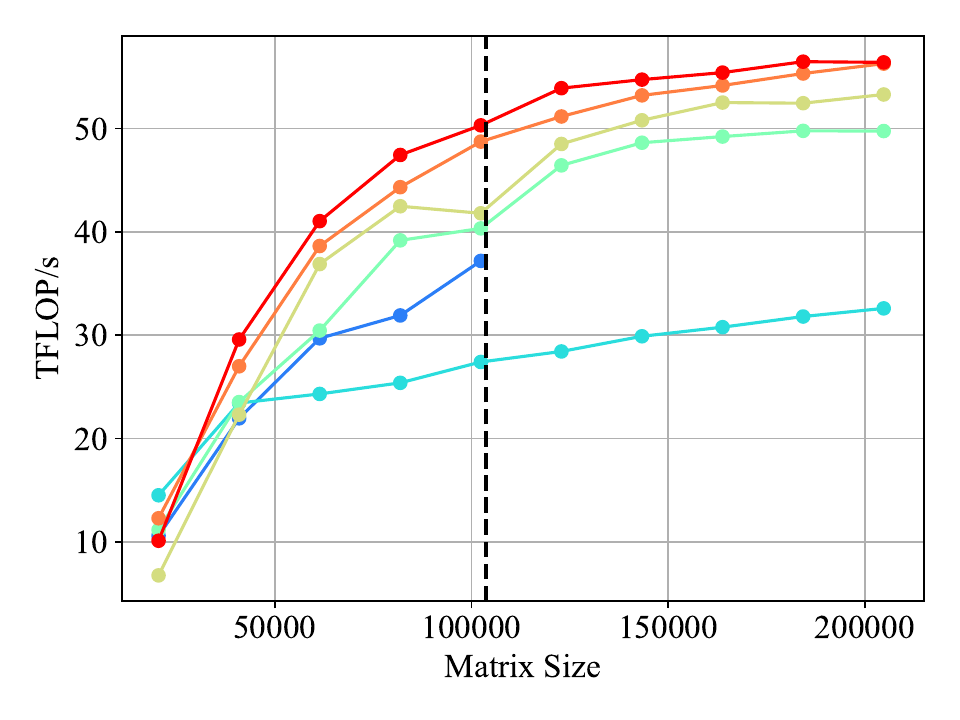}
    \caption{H100 with PCIe Gen5.}
\end{subfigure}
\hfill
\begin{subfigure}[b]{0.24\linewidth}
    \includegraphics[width=\linewidth]{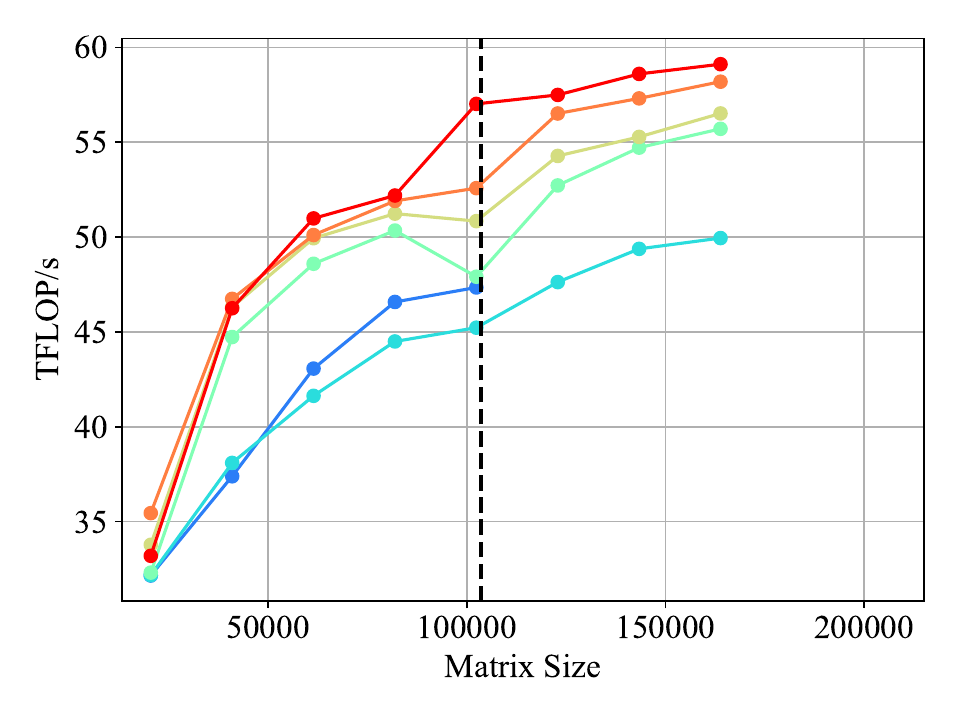}
    \caption{GH200 with NVLink-C2C.}
\end{subfigure}
\caption{Single GPU FP64 Cholesky performance on different GPUs w/ OOC support. Dashed line: 80GB GPU memory limit.}

\label{fig:single-FLops}
\end{figure*}

We compare the performance of our three different proposed implementations against cuSOLVER Cholesky factorization, which does not support OOC, and two implementations of the left-looking Cholesky factorization using the static scheduler of PLASMA~\cite{agullo2009numerical} supplemented by our naive OOC, referred to as \textit{sync} and \textit{async}. In the first variant, \textit{sync}, data is transferred from the CPU to the GPU just before computation, and the results are returned to the CPU afterward. To improve performance this naive OOC implementation, a multi-stream version of \textit{sync} is used, called \textit{async}, as described in Section~\ref{subsec:ooc}. We effectively overlap expensive data movement with useful computations by running data transfer and computation on multiple streams.

The results, as displayed in Figure~\ref{fig:single-FLops}, demonstrate that our optimized method achieves up to $16.1$ TFlop/s on the A100-PCIe, $54.7$ TFlop/s on the H100-PCIe, and 58.9 TFlop/s on the GH200-NVL-C2C using \textit{V3}. \textit{V1} and \textit{V2} score lower performance compared to \textit{V3}, but both outperform cuSOLVER, \textit{sync}, and \textit{async} implementations. 
These absolute performance numbers for \textit{V3} are within 95\% of the GEMM theoretical peak performance, which
highlights the effectiveness of our memory management strategies. An important observation from these experiments is the underperformance of \textit{async} implementation compared to \textit{V1}. This lesser efficiency of \textit{async} can be attributed primarily to the overhead associated with memory allocation (cudaMalloc) and deallocation (cudaFree) processes—such overheads impact performance, particularly in high-intensity computational environments.

\subsubsection{Event traces}
\begin{figure}[!ht]
\begin{subfigure}[b]{0.41\linewidth}
    \includegraphics[width=1\linewidth]{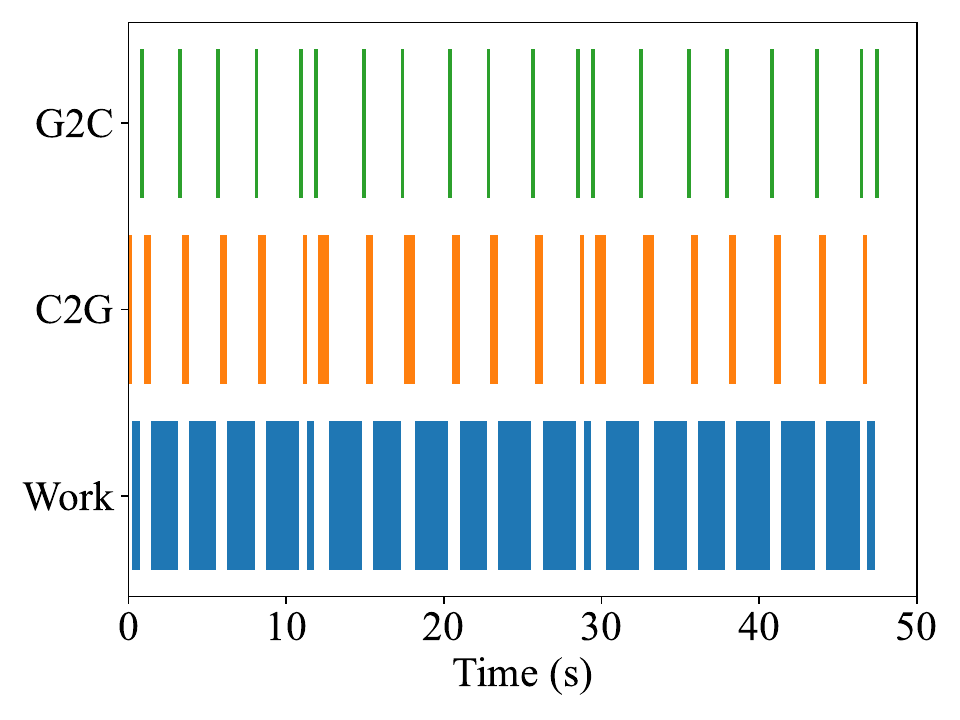}
    \caption{H100 w/ PCIe: sync.}
    \label{fig:event-traces-H100-sync}
\end{subfigure}
\hspace{2mm}
\begin{subfigure}[b]{0.41\linewidth}
    \includegraphics[width=\linewidth]{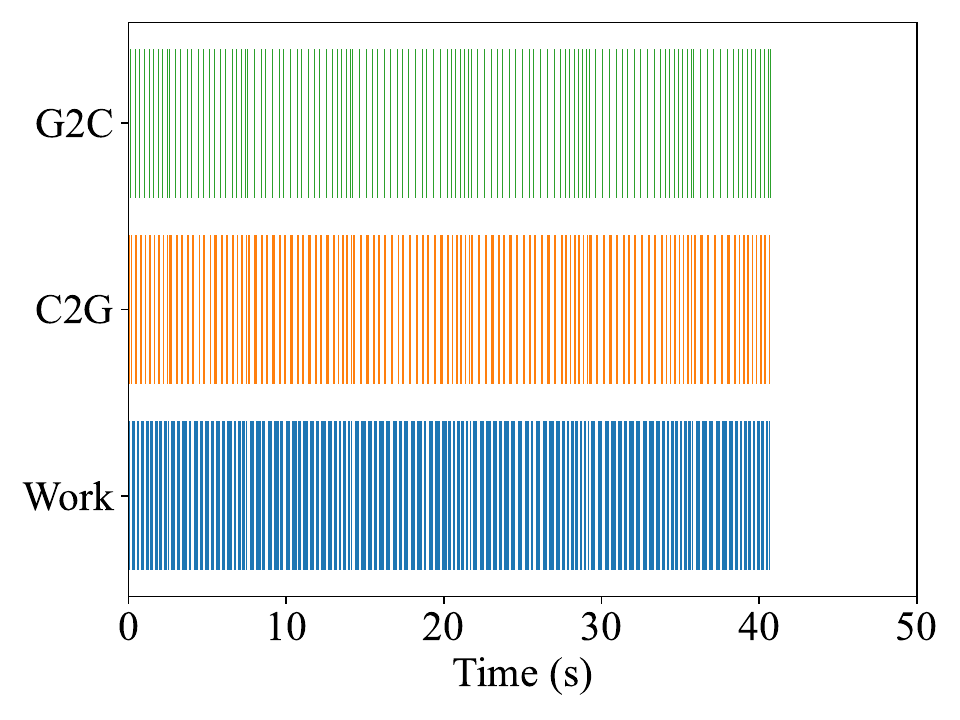}
    \caption{GH200 w/ C2C: sync.}
    \label{fig:event-traces-GH200-sync}
\end{subfigure}\hspace{2mm}
\begin{subfigure}[b]{0.41\linewidth}
    \includegraphics[width=\linewidth]{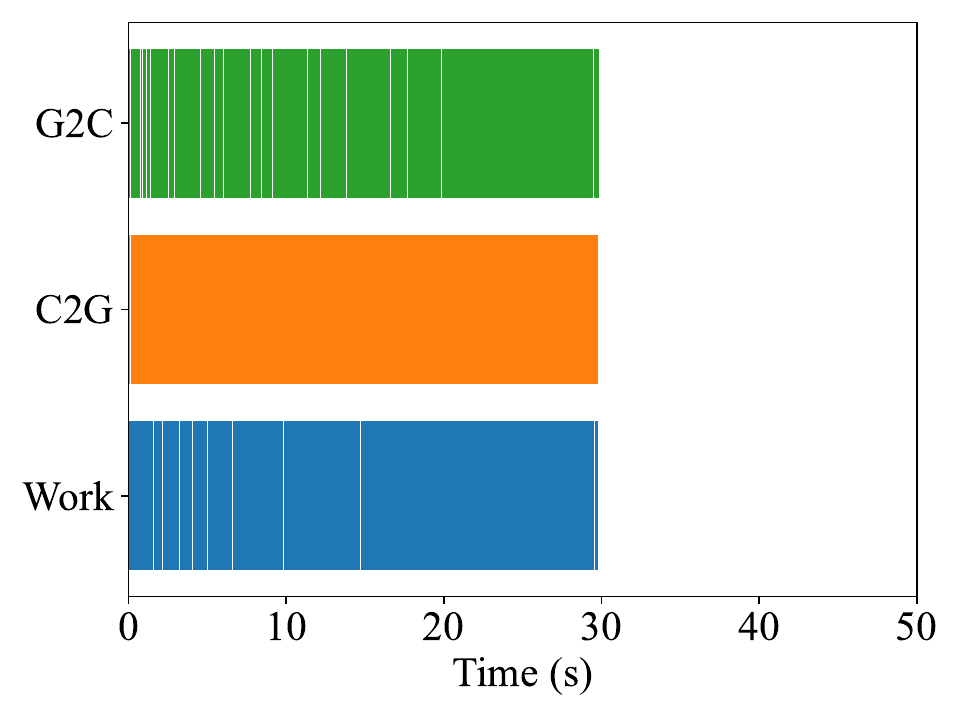}
    \caption{H100 w/ PCIe: async.}
    \label{fig:event-traces-H100-async}
\end{subfigure}
\hspace{2mm}
\begin{subfigure}[b]{0.41\linewidth}
    \includegraphics[width=\linewidth]{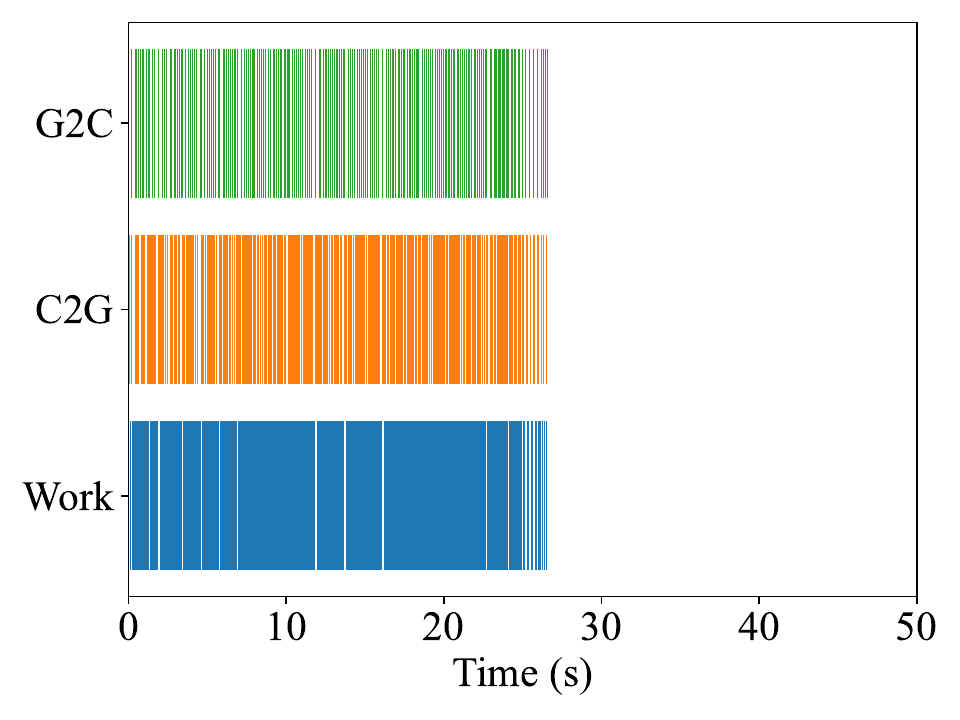}
    \caption{GH200 w/ C2C: async.}
    \label{fig:event-traces-GH200-async}
\end{subfigure}
\hfill
\begin{subfigure}[b]{0.41\linewidth}
    \includegraphics[width=\linewidth]{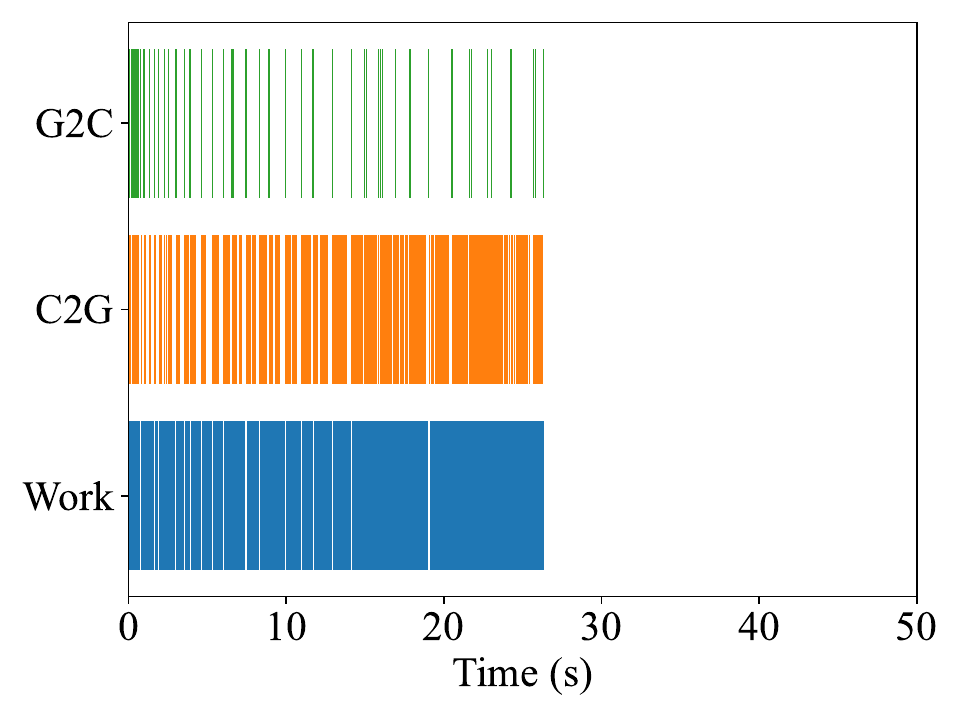}
    \caption{H100 w/ PCIe: V3.}
    \label{fig:event-traces-H100-V3}
\end{subfigure}
\hfill
\begin{subfigure}[b]{0.41\linewidth}
    \includegraphics[width=\linewidth]{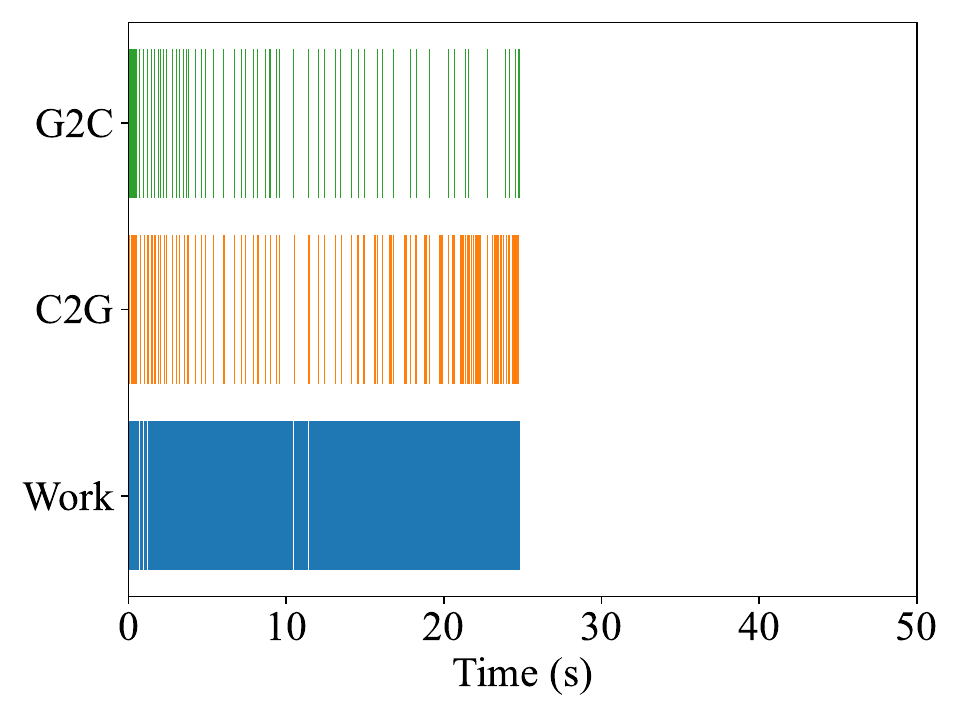}
    \caption{GH200 w/ C2C: V3.}
    \label{fig:event-traces-GH200-V3}
\end{subfigure}
\caption{Traces on a single GPU w/ matrix size 160K x 160K.  Memory transfers from the GPU to the CPU are indicated in \textcolor{green}{green} on the `C2G' row. Transfers from the CPU to the GPU are marked in \textcolor{orange}{orange} on the '
`G2C' row. Computational tasks are represented in \textcolor{blue}{blue} on the `Work' row.}
\label{fig:event-traces}
\end{figure}

To visualize the advantage of our proposed implementation, we present the traces of data movement and computation using a single GPU in Figure~\ref{fig:event-traces}. The experiment is conducted on H100-PCIe and GH200-NVL-C2C servers, focusing on matrices with 160k x 160k dimensions. In the figure, memory transfers from GPU to CPU are highlighted in \textcolor{green}{green} on the `C2G' row. Conversely, memory transfers from CPU to GPU are shown in \textcolor{orange}{orange} on the `G2C' row. Computational tasks are depicted in \textcolor{blue}{blue} on the 
`Work' row.

A closer examination of Figures~\ref{fig:event-traces-H100-sync} and~\ref{fig:event-traces-GH200-sync} reveals idle periods within the computation process, indicating times when the GPU awaits data transfer. In contrast, as illustrated in Figure~\ref{fig:event-traces-H100-async} and~\ref{fig:event-traces-GH200-async}, there is an overlap of computation with data movement, showcasing efficient parallelism. Another interesting observation in Figures~\ref{fig:event-traces-H100-V3} and~\ref{fig:event-traces-GH200-V3} is the reduction in CPU to GPU data copying events. This efficiency is attributed to implementing the data caching progress table, which effectively keeps the data alive on GPU memory and, therefore, decreases the volume of data movement. Moreover, the H100-PCIe server tends to favor using larger data tiles than the GH200-NVL-C2C. This preference of GH200-NVL-C2C for smaller tiles is due to the interconnect high bandwidth. However, on the H100-PCIe server, the primary constraint is data movement due to a slow PCIe interconnect, making larger tiles a more suitable choice in this context.

\subsubsection{Data Movement Volume on a Single GPU}

\begin{figure*}[!ht]
\begin{subfigure}[b]{0.25\linewidth}
    \includegraphics[width=\textwidth]{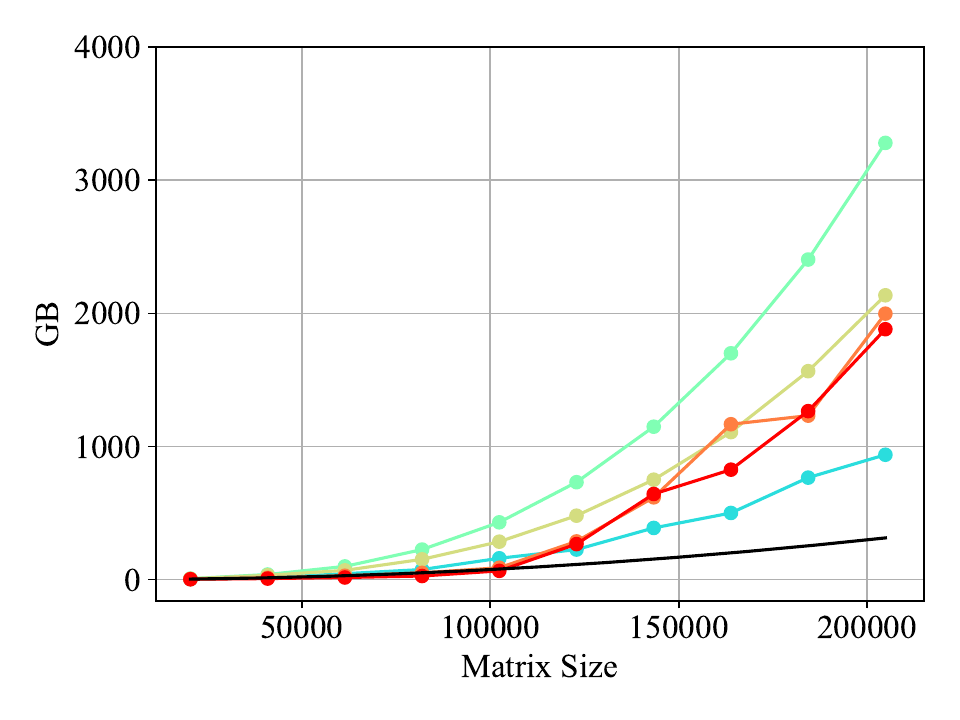}
    \caption{A100\_PCIE\_C2G}
\end{subfigure}
\hfill
\begin{subfigure}[b]{0.25\linewidth}
    \includegraphics[width=\textwidth]{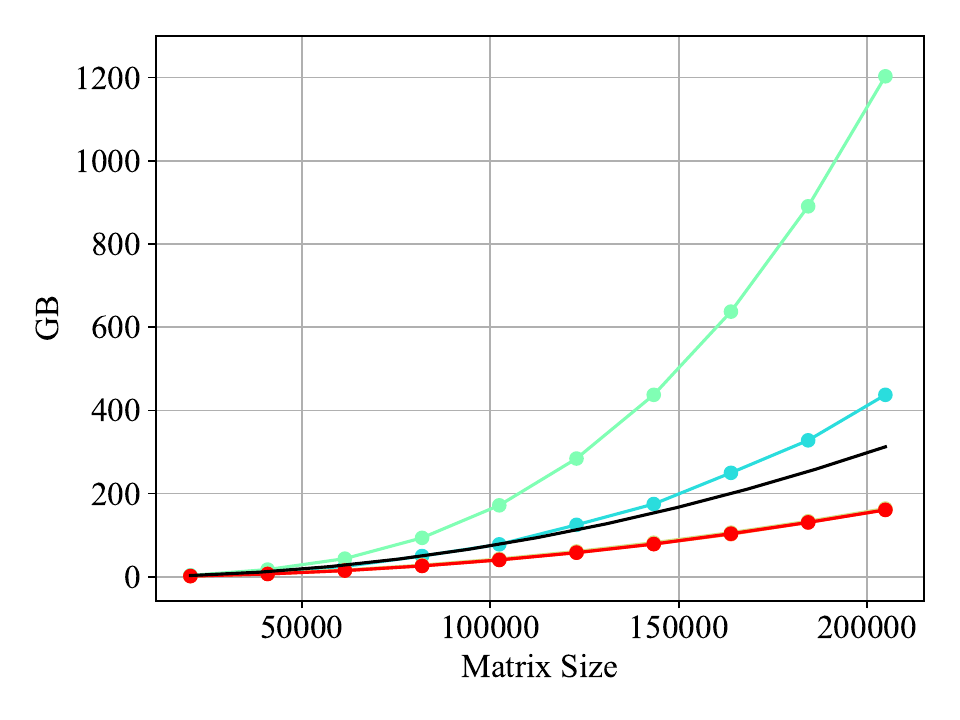}
    \caption{A100\_PCIE\_G2C}
\end{subfigure}
\hfill
\begin{subfigure}[b]{0.25\linewidth}
    \includegraphics[width=\textwidth]{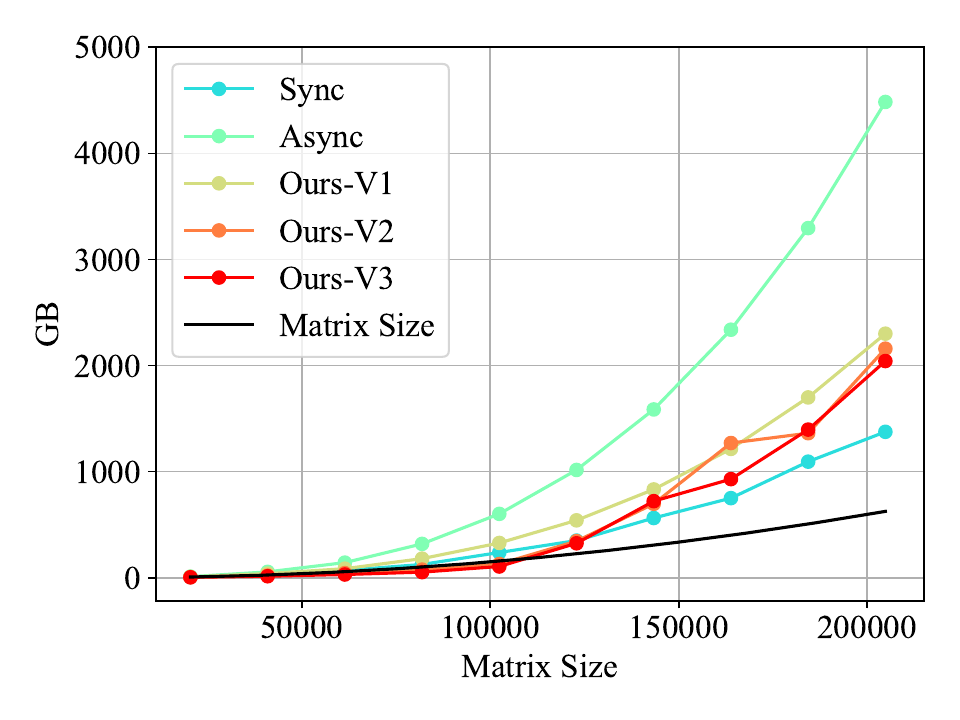}
    \caption{A100\_PCIE\_Total}
\end{subfigure}
\hfill
\begin{subfigure}[b]{0.25\linewidth}
    \includegraphics[width=\textwidth]{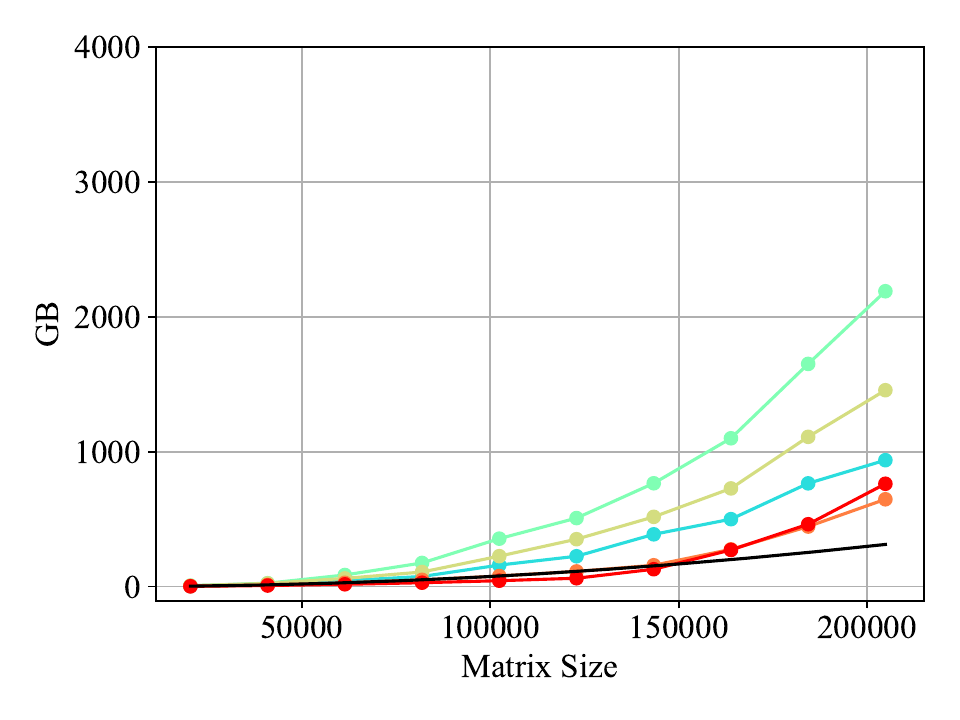}
    \caption{H100\_PCIE\_C2G}
\end{subfigure}
\hfill
\begin{subfigure}[b]{0.25\linewidth}
    \includegraphics[width=\textwidth]{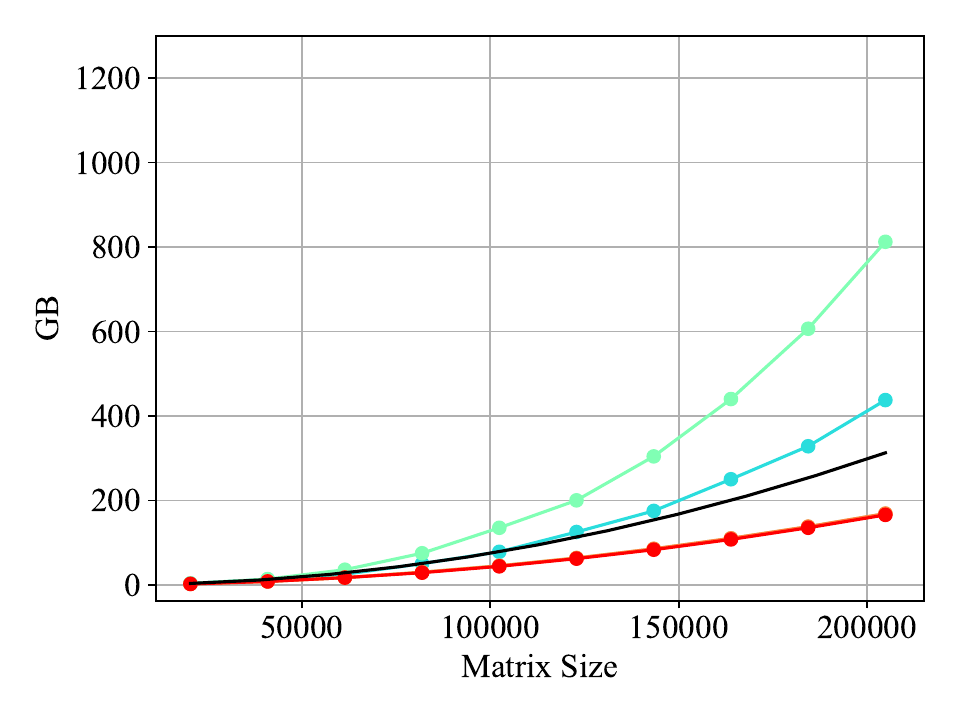}
    \caption{H100\_PCIE\_G2C}
\end{subfigure}
\hfill
\begin{subfigure}[b]{0.25\linewidth}
    \includegraphics[width=\textwidth]{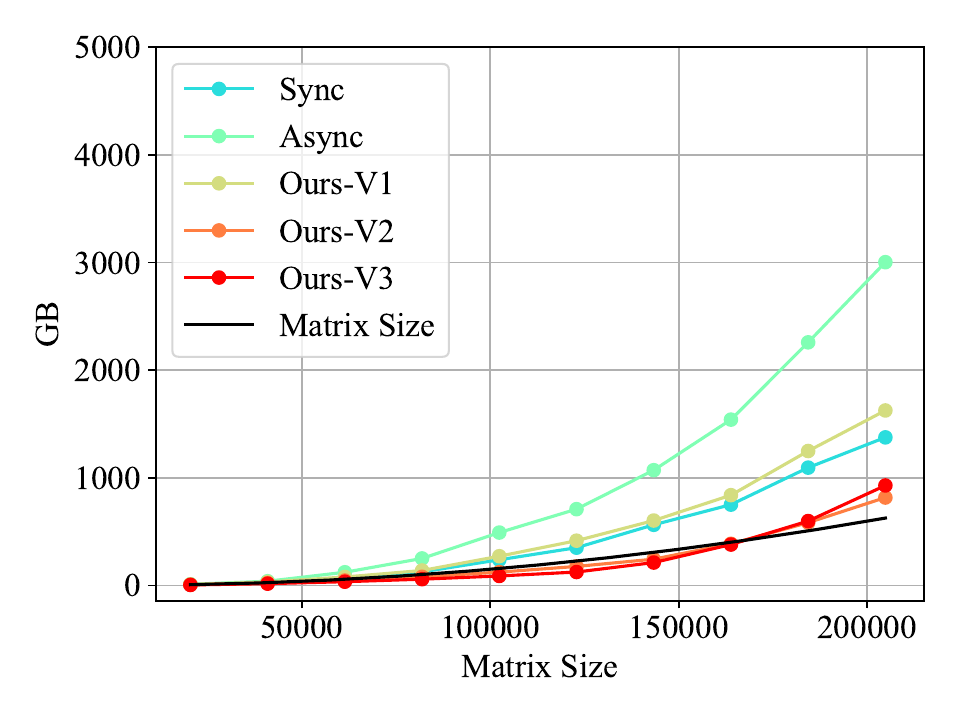}
    \caption{H100\_PCIE\_Total}
\end{subfigure}
\hfill
\begin{subfigure}[b]{0.25\linewidth}
    \includegraphics[width=\textwidth]{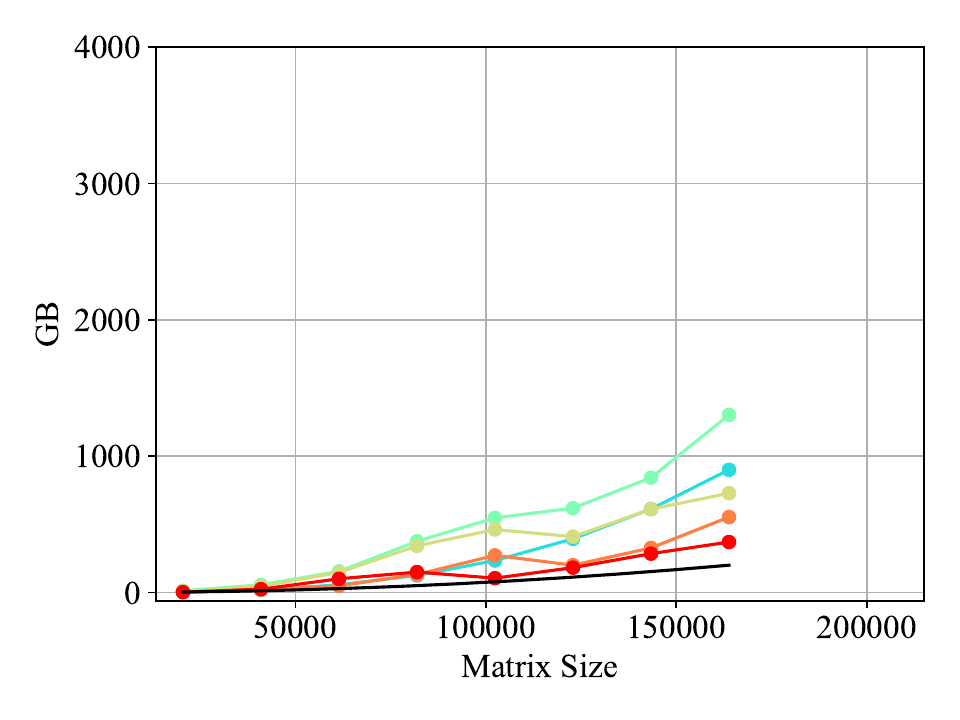}
    \caption{GH200\_NVL\_C2G}
\end{subfigure}
\hfill
\begin{subfigure}[b]{0.25\linewidth}
    \includegraphics[width=\textwidth]{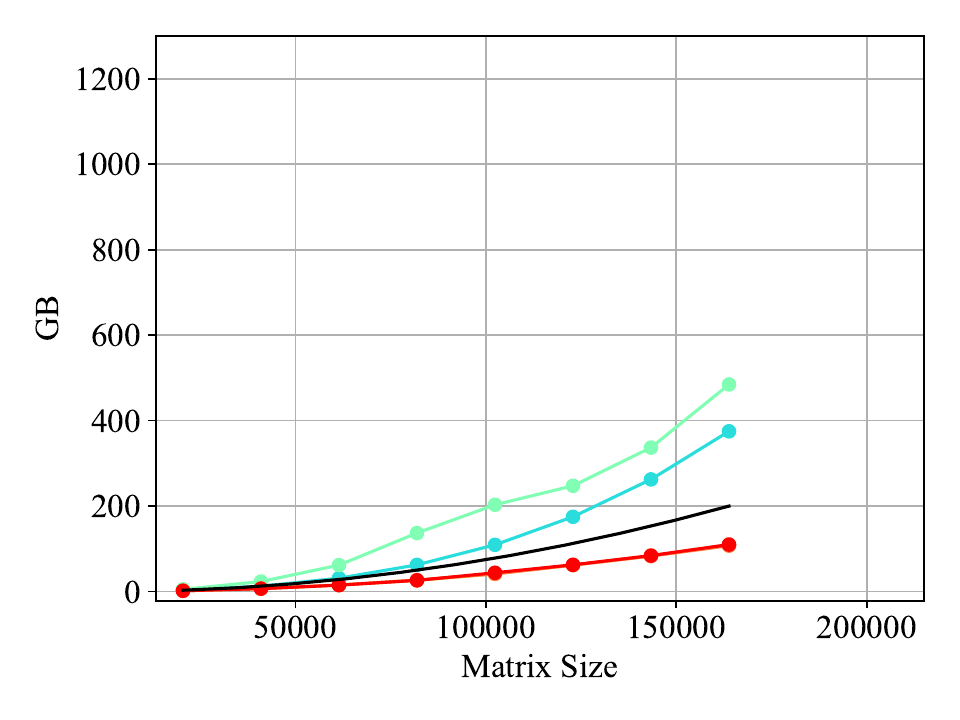}
    \caption{GH200\_NVL\_G2C}
\end{subfigure}
\hfill
\begin{subfigure}[b]{0.25\linewidth}
    \includegraphics[width=\textwidth]{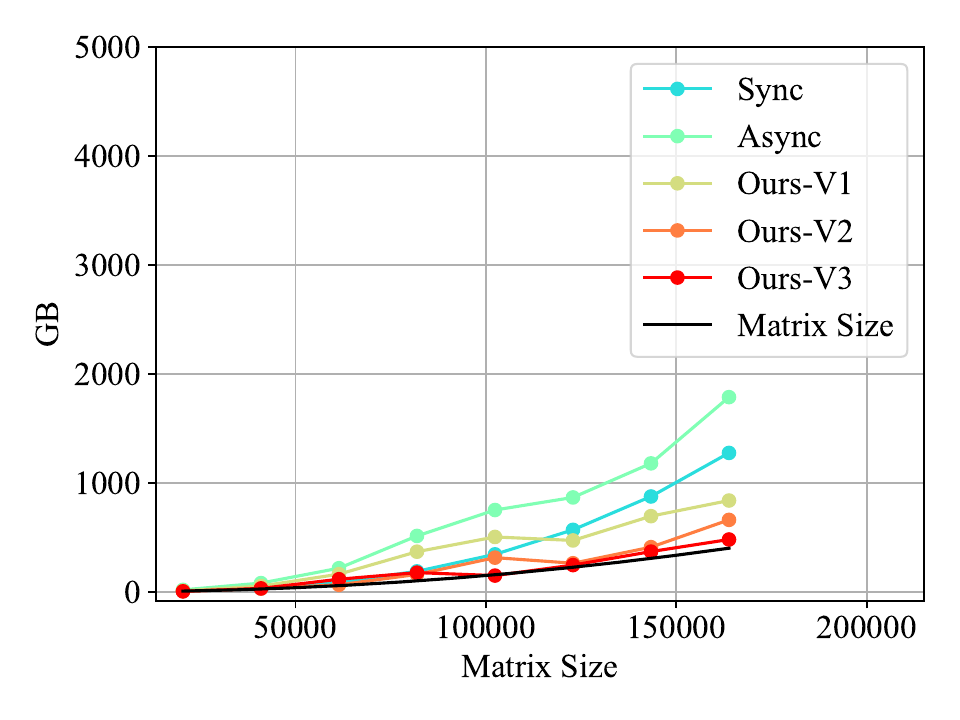}
    \caption{GH200\_NVL\_TOTAL}
\end{subfigure}
\hfill
\caption{Volume of data communication (C2G, G2C, and Total) across different implementations on single GPU.}
\label{fig:volume-communication}
\end{figure*}

We measure the volume of data movement on a single GPU setup using cuSOLVER and our various implementations on the A100-PCIe, H100-PCIe, and GH200 Grace Hopper Superchip. As shown in Figure~\ref{fig:volume-communication}, the cuSOLVER data movement volume is the same as the matrix size, and the data movement volume of \textit{Async} is the highest. The data movement volume from the GPU to CPU of our V1-V3 approaches is almost half of the matrix size, because we only copy the triangular part of the symmetric matrix. The data movement volume of \textit{Sync} is smaller than that of \textit{Async} and sometimes smaller than that of our V1-V3 techniques because it uses larger tiles, and the tile number is smaller, which makes the data movement volume smaller, which ultimately impacts performance. The volume of data movement of a 160k$\times$160k matrix for our method is reduced to $935.7$GB on the A100-PCIe, $437.4$GB on the H100-PCIe, and $480$GB on the GH200, indicating more efficient data handling, up to 3X less memory than the \textit{async} approach. Generally, the data movement volume relation is \textit{V3} $<$ \textit{V2} $<$ \textit{V1} $<$ \textit{Async}.  The volume of data varies between GPUs because we tune the tile size for optimal performance on each GPU, implementation, and for each matrix size.


\subsection{Multi-GPU Performance}


We conduct an additional series of experiments to evaluate the performance of the proposed implementation across multiple GPUs, using configurations ranging from one to four units.  The comparison focuses on FP64 computations.

\subsubsection{Multi-GPU Flop/s Comparison}
We compared the flops/s performance of the proposed implementations across different accelerators: A100-PCIe, H100-PCIe, and the GH200 Grace Hopper Superchip. The A100-PCIe and H100-PCIe servers have a single CPU connected to multiple GPUs via the PCIe bus. 

\begin{figure*}[!ht]
\begin{subfigure}[b]{0.25\linewidth}
    \includegraphics[width=\linewidth]{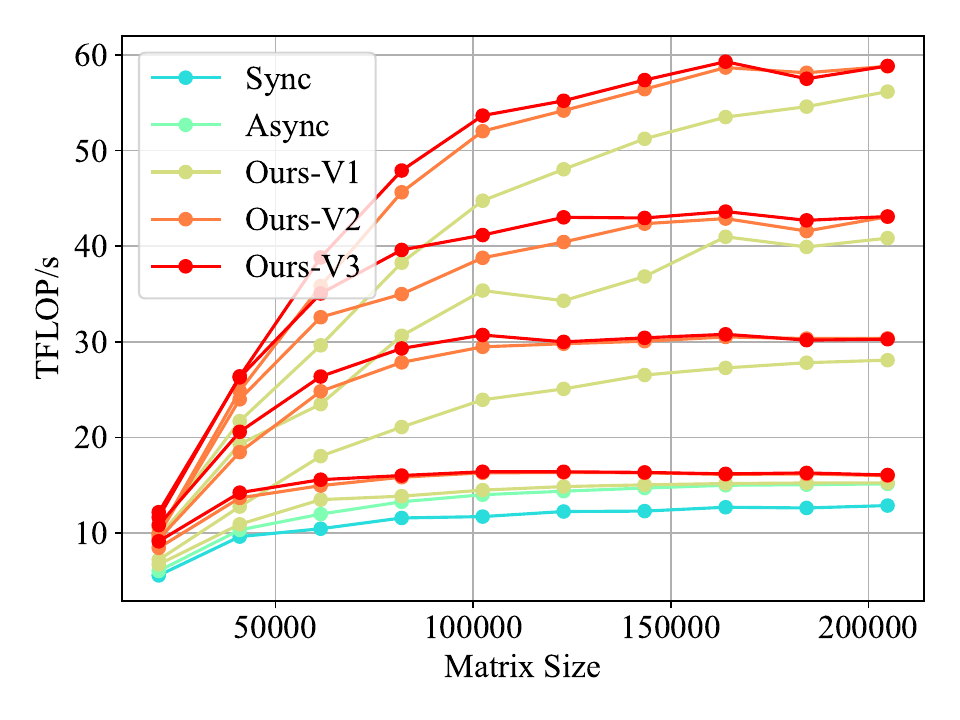}
    \caption{A100 with PCIe Gen4.}
\end{subfigure}
\hfill
\begin{subfigure}[b]{0.25\linewidth}
    \includegraphics[width=\linewidth]{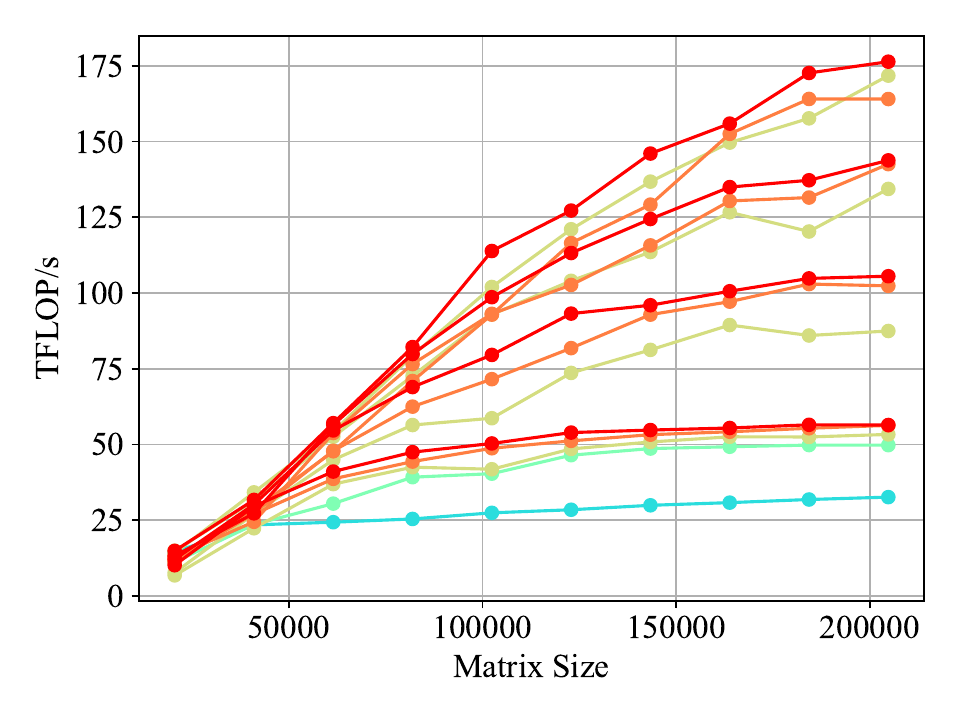}
    \caption{H100 with PCIe Gen5.}
\end{subfigure}
\hfill
\begin{subfigure}[b]{0.25\linewidth}
    \includegraphics[width=\linewidth]{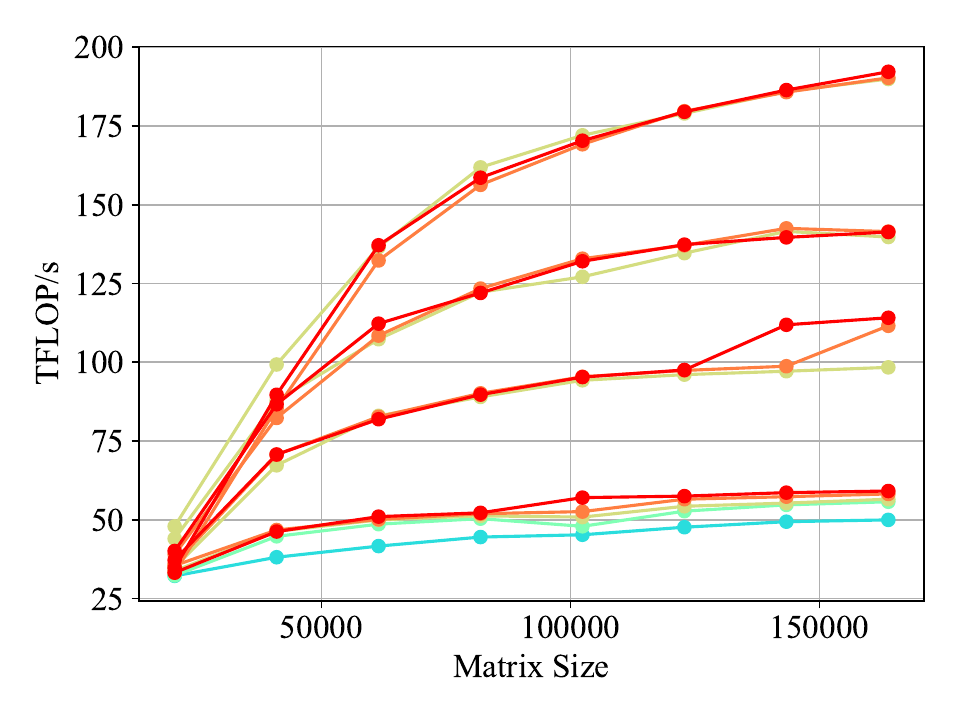}
    \caption{GH200 with NVLink-C2C.}
\end{subfigure}
\caption{Multi-GPU FP64 Cholesky performance with OOC support on A100, H100, and GH200.
}
\label{fig:three_versions_multiple}
\end{figure*}

Our results from the multi-GPU setup shown in Figure~\ref{fig:three_versions_multiple} indicate remarkable performance on four GPUs: achieving $59.3$ TFlop/s on the A100-PCIe, $176.3$ TFlop/s on the H100-PCIe, and $185.5$ TFlop/s on the GH200 system. As shown in the scalability figure, each version has four curves representing the use of 4 GPUs, 3 GPUs, 2 GPUs, and 1 GPU, with performance increasing as the number of GPUs increases.

It is important to note a distinct trend observed in the performance growth between the H100-PCIe and GH200-NVL-C2C accelerators. Specifically, the H100-PCIe demonstrate a lower slope for the rate of execution relative to the GH200-NVL-C2C as the matrix size increases. This trend is primarily attributed to the bottleneck caused by data movement speed, which lags significantly behind the computation speed in the case of the H100-PCIe server. All in all, NVLink-C2C interconnect delivers high bandwidth and can sustain high demands of data motion, when it comes to FPP64 computations.

\begin{figure*}[!ht]
\captionsetup[subfigure]{justification=centering}
\begin{subfigure}[b]{0.25\linewidth}
\includegraphics[width=\linewidth]{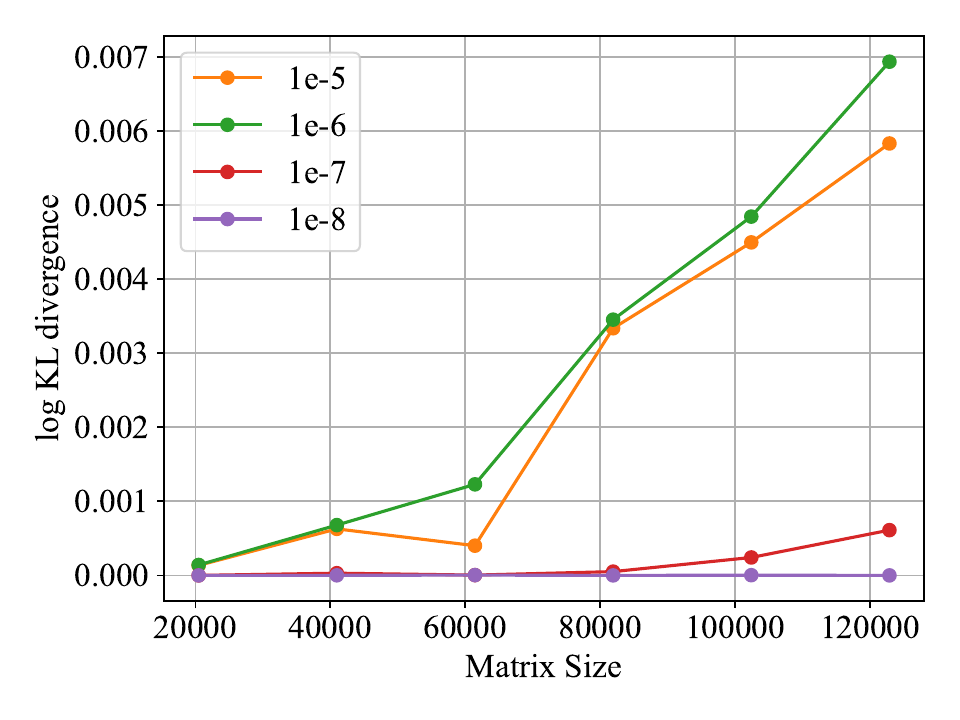}
\caption{Weak Correlation \\ $\boldsymbol{\theta}= (1, 0.02627, 0.5)$ }
\end{subfigure}
\hfill
\begin{subfigure}[b]{0.25\linewidth}
\includegraphics[width=\linewidth]{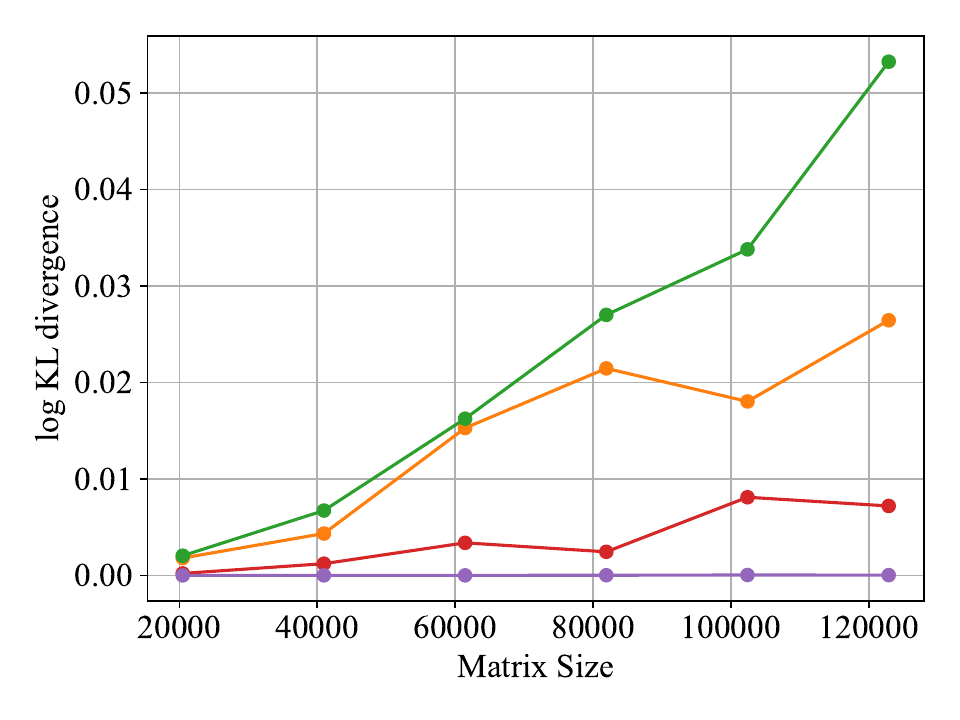}
\caption{Medium Correlation \\ $\boldsymbol{\theta}= (1, 0.078809, 0.5)$ }
\end{subfigure}
\hfill
\begin{subfigure}[b]{0.25\linewidth}
\includegraphics[width=\linewidth]{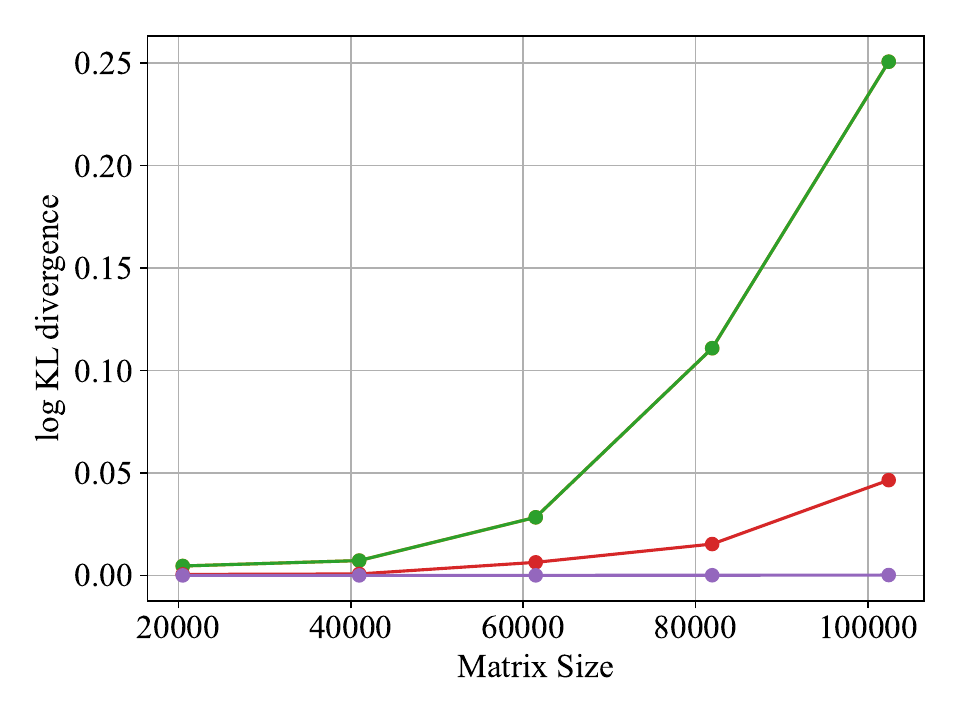}
\caption{Strong Correlation \\ $\boldsymbol{\theta}= (1, 0.210158, 0.5)$ }
\end{subfigure}
\caption{KL divergence for varying matrix sizes (y-axis: log10 scale). Spatial correlation decreases left to right with $\beta$ values $0.02627, 0.078809, and 0.210158$. The legend shows MxP accuracy levels (smaller is better). }
\label{fig:KL-divergence}
\end{figure*}
\subsection{Mixed-Precision Assessment on GH200}
In this section, we evaluate the accuracy and performance of the left-looking algorithm using four different precision levels: FP8, FP16, FP32, and FP64. Given that the distribution of precisions across the specified tiles varies with different matrix data, we conduct the experiments through a geospatial data modeling application, as described in Section III (D). We consider three scenarios that may arise in such geospatial applications: weak, medium, and strong correlated data, which are discussed in detail below.

\subsubsection{Mixed-Precision Accuracy Assessment on GH200}

We rely on the KL divergence metric presented in Eq.~\ref{eq:kl-gaussian} to assess the accuracy of our MxP algorithm. The KL divergence metric can estimate the error introduced by applying approximation techniques to the likelihood function in equation~\ref{eq:likeli}. This is done by measuring the divergence between the estimated probability and original distributions. In this context, we evaluate the probability distribution of the MxP version of the likelihood function compared to the FP64 variant.

Figure~\ref{fig:KL-divergence} shows the KL divergence at different correlation levels ($\beta$). Specifically, weak correlation is represented by $\beta = 0.02627$, medium correlation by $\beta = 0.078809$, and strong correlation by $\beta = 0.210158$. A key aspect of this analysis is the relationship between $\beta$ and the precision of the tiles: lower values of $\beta$ lead to higher use of low-precision tiles.

Based on the data shown in Figure~\ref{fig:KL-divergence}, for weak correlation ($\beta = 0.02627$), the KL-divergence of the MxP approach compared to the FP64 variant is relatively small. This improves further with a higher accuracy threshold (e.g., $1e-8$). As the correlation increases (from the left subfigure to the right), the KL divergence also increases, requiring a higher accuracy threshold for the MxP distribution to closely resemble the original one. Generally, while there is no strict threshold for KL divergence that significantly impacts the modeling process, smaller values are preferred.

\subsubsection{Mixed-Precision Performance Comparison on GH200}

\begin{figure*}[!ht]
\captionsetup[subfigure]{justification=centering}
\begin{subfigure}[b]{0.27\linewidth}
\includegraphics[width=\linewidth]{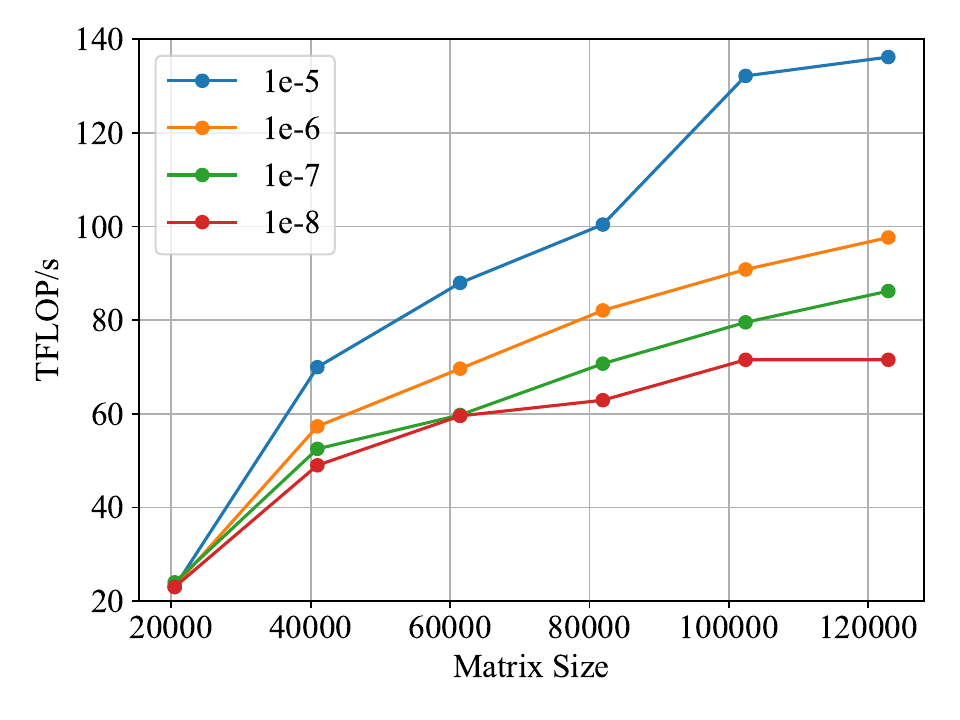}
\caption{Weak Correlation \\ $\boldsymbol{\theta}= (1, 0.02627, 0.5)$ }
\label{fig:MxP-TFLOPS-weak}
\end{subfigure}
\hfill
\begin{subfigure}[b]{0.27\linewidth}
\includegraphics[width=\linewidth]{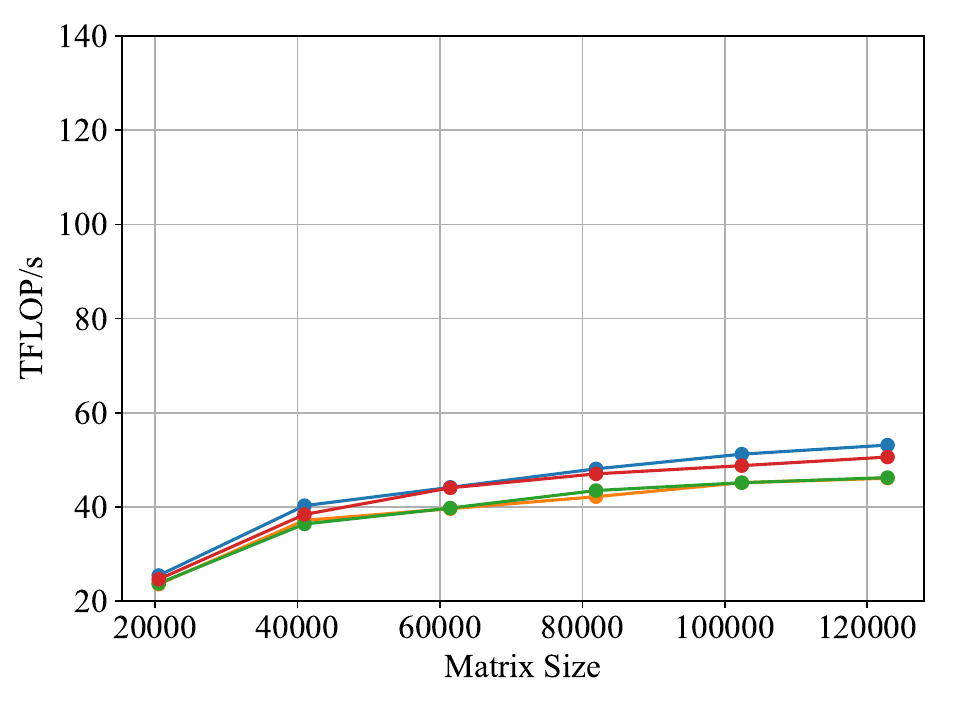}
\caption{Medium Correlation \\ $\boldsymbol{\theta}= (1, 0.078809, 0.5)$ }
\label{fig:MxP-TFLOPS-medium}
\end{subfigure}
\hfill
\begin{subfigure}[b]{0.27\linewidth}
\includegraphics[width=\linewidth]{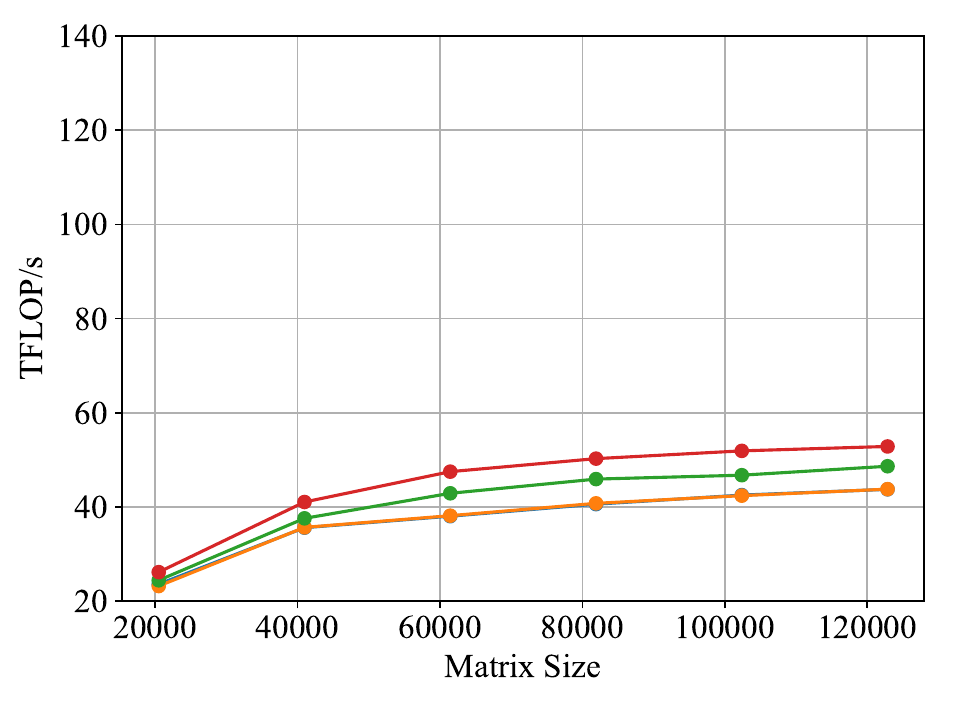}
\caption{Strong Correlation \\ $\boldsymbol{\theta}= (1, 0.210158, 0.5)$ }
\label{fig:MxP-TFLOPS-strong}
\end{subfigure}
\caption{Performance on single GH200 GPU with MxP calculations using different matrix sizes and accuracy levels. 
}
\label{fig:MxP-TFLOPS}
\end{figure*}

We now assess the performance of the MxP algorithm using the same settings of the aforementioned subsection. Figure~\ref{fig:MxP-TFLOPS} provides a detailed analysis of the MxP factorization performance in our target geospatial data application. The legend indicates the accuracy levels, where lower accuracy introduces more low-precision tiles, and higher accuracy incorporates more high-precision tiles.
The reported performance results are based on a single GH200 GPU.

Focusing on the results in Figure~\ref{fig:MxP-TFLOPS-weak}, the MxP method achieved up to $136.1$ TFlop/s, using the lowest accuracy level of $\beta = 10^{-5}$. This performance benchmark is particularly significant as it reflects the effective use of precision in computational processes. Performance declines with higher accuracy due to the reliance on higher precision computation and data movement. We also observed that the demand for higher precision tiles rises as spatial location correlations increase. Consequently, Figures~\ref{fig:MxP-TFLOPS-medium} and~\ref{fig:MxP-TFLOPS-strong} demonstrate a significant decrease in performance due to the increased utilization of higher precision tiles.

Another observation is that with strong correlation, higher precision tiles are necessary, which allows higher accuracy (1e-8) to outperform lower accuracy ($10^{-5}$). This improvement occurs because the additional overhead associated with data up/down-casting for FP32 tiles impacts the overall performance. Nonetheless, using low-precision tiles offers advantages, especially regarding memory usage savings, highlighting their importance in specific computational contexts.

\begin{figure*}[!ht]
\captionsetup[subfigure]{justification=centering}
\begin{subfigure}[b]{0.27\linewidth}
\includegraphics[width=\linewidth]{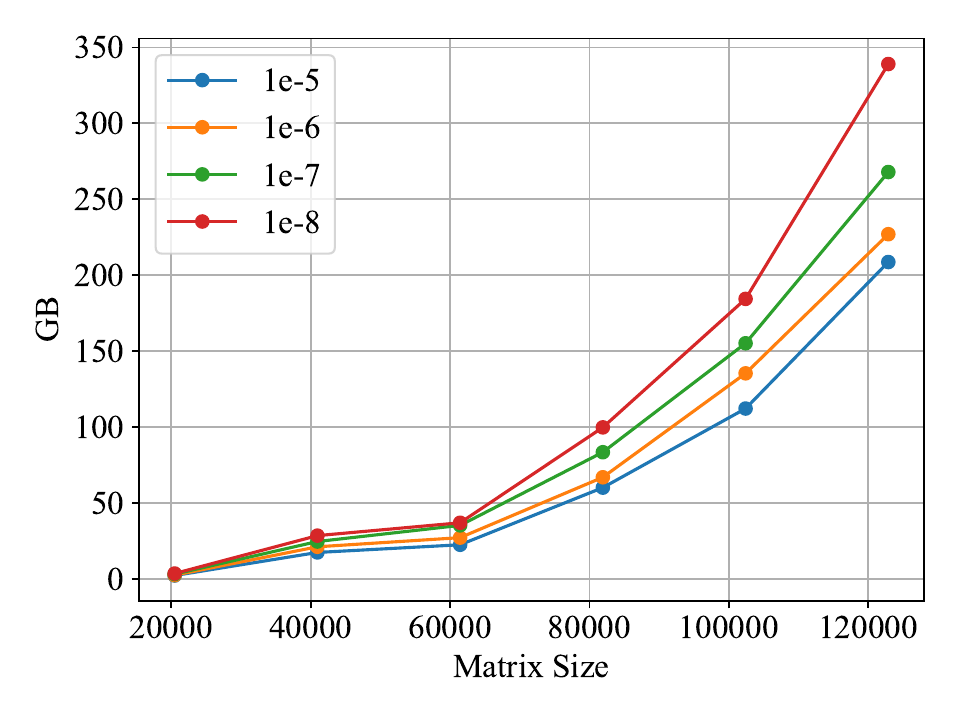}
\caption{Weak Correlation \\ $\boldsymbol{\theta}= (1, 0.02627, 0.5)$ }
\end{subfigure}
\hfill
\begin{subfigure}[b]{0.27\linewidth}
\includegraphics[width=\linewidth]{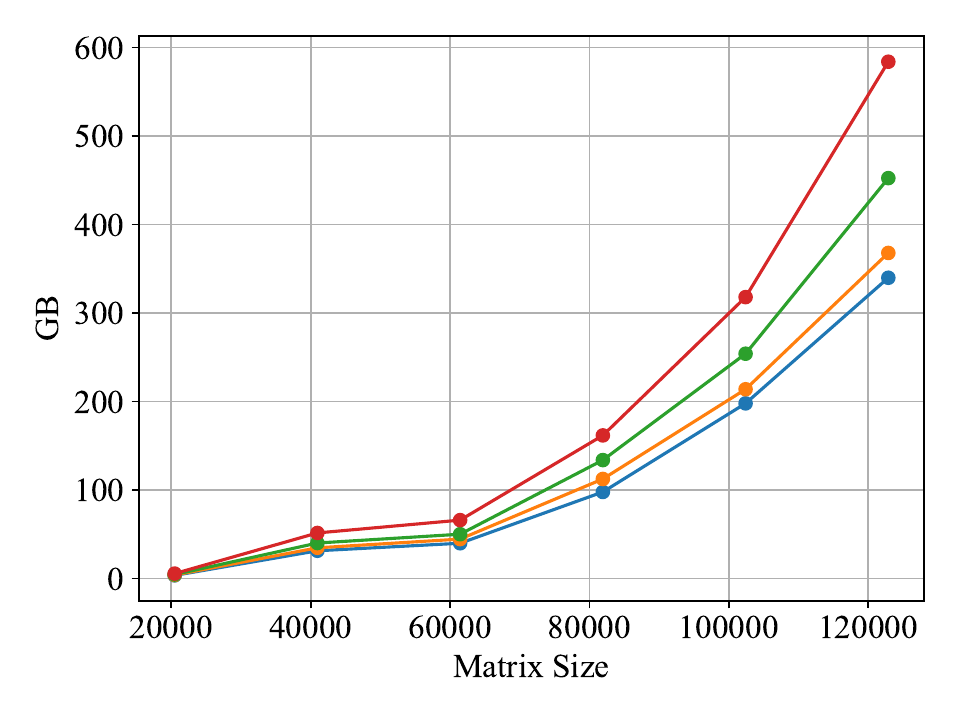}
\caption{Medium Correlation \\ $\boldsymbol{\theta}= (1, 0.078809, 0.5)$ }
\end{subfigure}
\hfill
\begin{subfigure}[b]{0.27\linewidth}
\includegraphics[width=\linewidth]{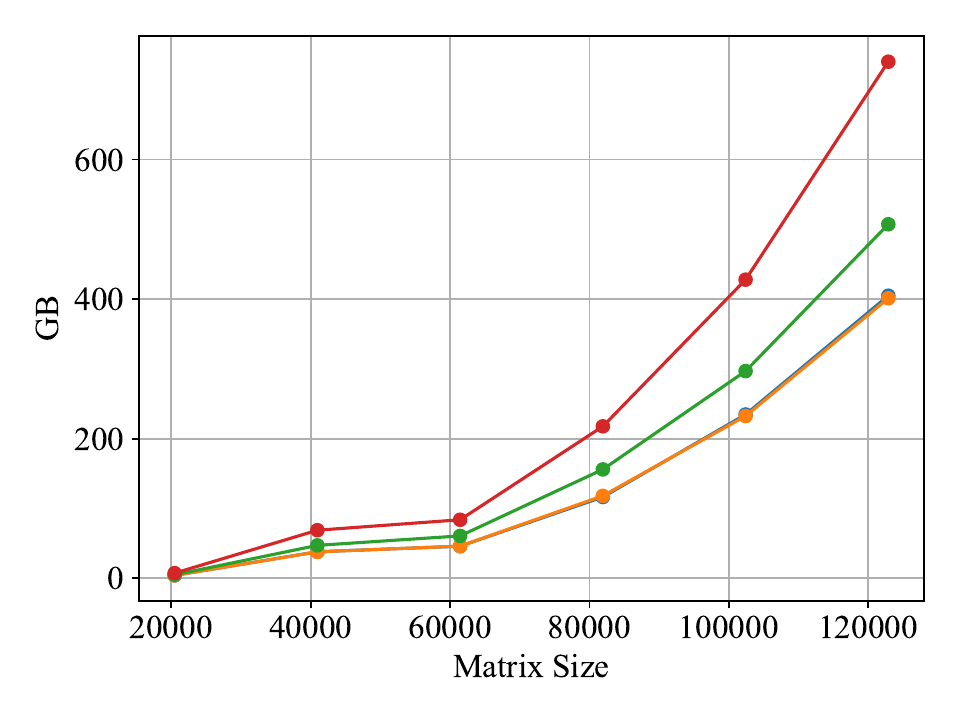}
\caption{Strong Correlation \\ $\boldsymbol{\theta}= (1, 0.210158, 0.5)$ }
\end{subfigure}
\caption{Data Movement volume with MxP calculations using different matrix sizes and accuracy levels.
}
\label{fig:MxP-volume}
\end{figure*}

\begin{figure*}[!ht]
\captionsetup[subfigure]{justification=centering}
\begin{subfigure}[b]{0.25\linewidth}
\includegraphics[width=\linewidth]{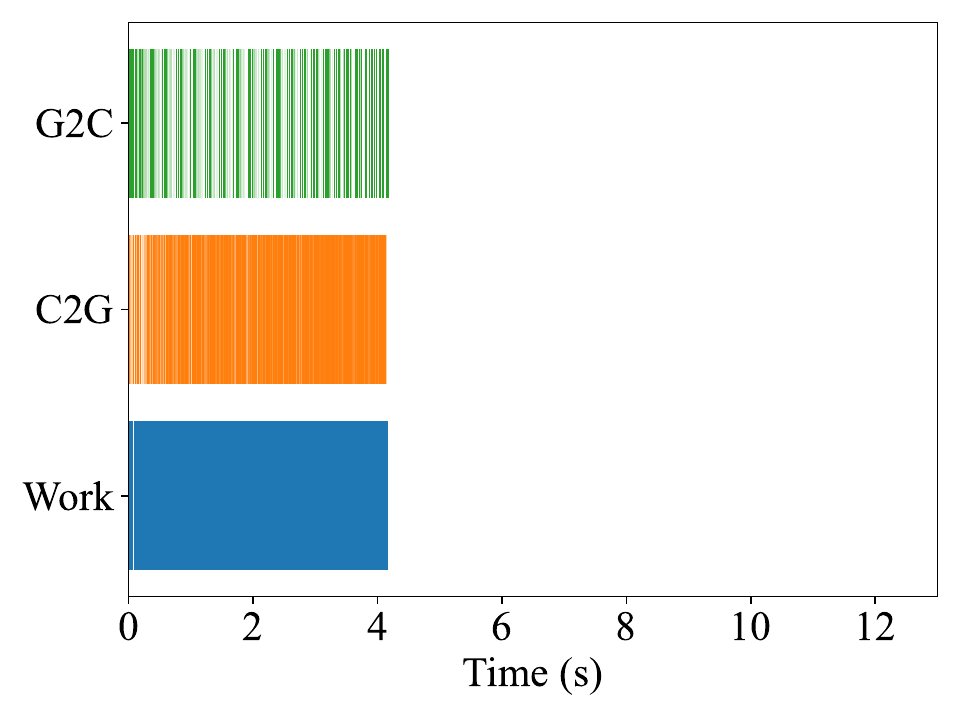}
\caption{Weak Correlation \\ $\boldsymbol{\theta}= (1, 0.02627, 0.5)$ }
\end{subfigure}
\hfill
\begin{subfigure}[b]{0.25\linewidth}
\includegraphics[width=\linewidth]{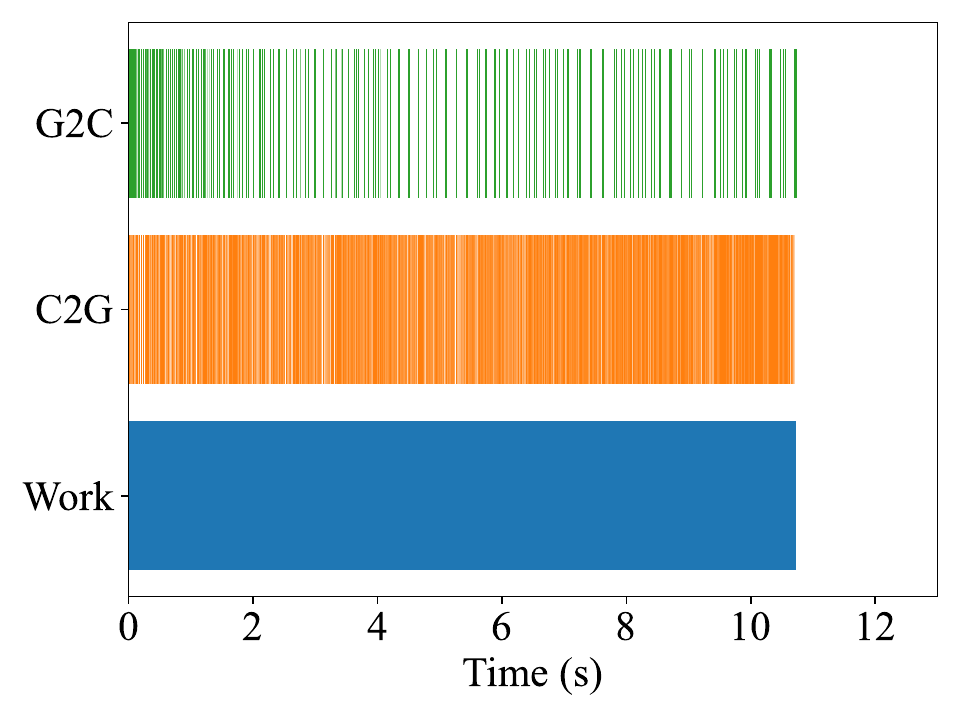}
\caption{Medium Correlation \\ $\boldsymbol{\theta}= (1, 0.078809, 0.5)$ }
\end{subfigure}
\hfill
\begin{subfigure}[b]{0.25\linewidth}
\includegraphics[width=\linewidth]{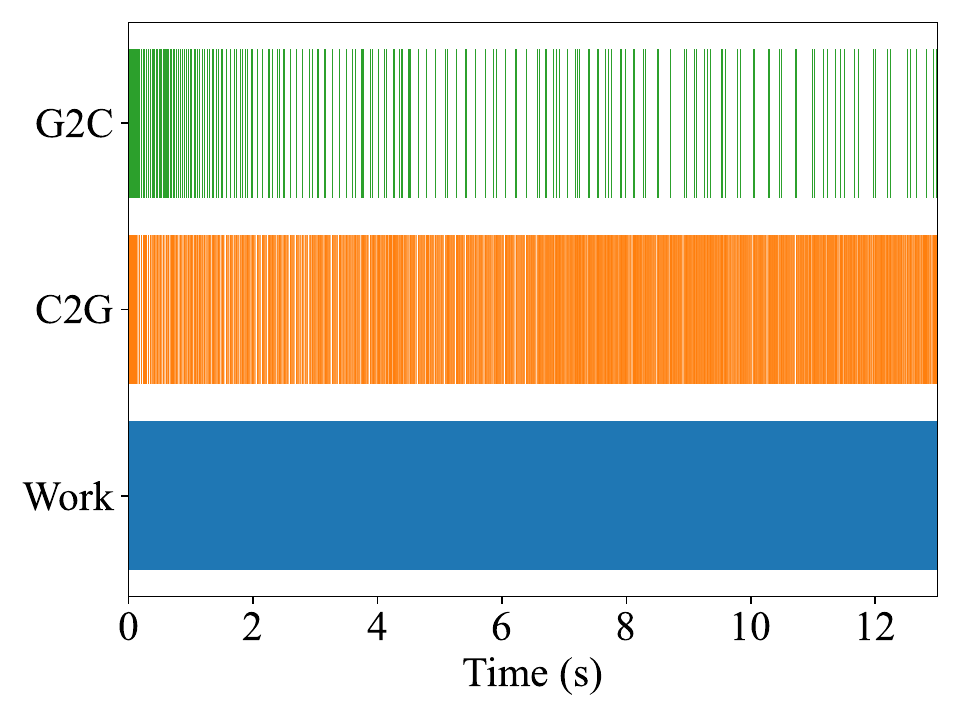}
\caption{Strong Correlation \\ $\boldsymbol{\theta}= (1, 0.210158, 0.5)$ }
\end{subfigure}
\caption{Traces on a single GH200 GPU w/ matrix size 100k$\times$100k and different spatial data correlation levels.}
\label{fig:MxP-events}
\end{figure*}

In Figure~\ref{fig:MxP-volume}, we explore the data movement volume associated with different spatial correlation levels. Here, it is evident that high accuracy $10^{-8}$ results in the highest data movement volume, correlating with its usage of the highest precision tiles. Conversely, low accuracy $10^{-5}$ exhibits the least volume, indicative of lower precision tiles usage.

Further insights are provided in Figure~\ref{fig:MxP-events}, where we display event traces for different correlation levels while maintaining $10^{-5}$ accuracy. The matrix size for this part of the study is 100k$\times$100k. A key takeaway from this figure is the substantial reduction in computation time achieved by incorporating low-precision tiles, particularly in scenarios characterized by weak correlation. This finding highlights the impact of precision selection on computational efficiency, especially in large-scale spatial statistics applications.

Looking at the performance and trace results from the perspective of NVLink-C2C interconnect, this timely technology does not blink an eye when overwhelmed with high latency messages due to low precision arithmetic. The surface-to-volume remains the principal knob in ensuring the device does not run out of work while still delivering on its promises.


\section{Conclusion}
\label{sec:summary}
Out-of-device memory computations in high-performance computing are optimized utilizing NVLink on the Grace Hopper Superchip. Copying data asynchronously alongside device computations significantly reduces data transfer latencies. This is particularly beneficial for the MxP left-looking Cholesky factorization powered by an efficient static task scheduler, highlighting the potential for handling datasets beyond device memory limitations.

These findings showcase the capabilities of modern interconnect technologies in managing extensive data sets effectively. The methodologies developed in this study hold promise for further applications and optimizations in diverse computational scenarios, paving the way for future innovations in efficiently processing large-scale data.

\bibliographystyle{IEEEtran}
\bibliography{references} %

\begin{thebibliography}{10}
\providecommand{\url}[1]{#1}
\csname url@samestyle\endcsname
\providecommand{\newblock}{\relax}
\providecommand{\bibinfo}[2]{#2}
\providecommand{\BIBentrySTDinterwordspacing}{\spaceskip=0pt\relax}
\providecommand{\BIBentryALTinterwordstretchfactor}{4}
\providecommand{\BIBentryALTinterwordspacing}{\spaceskip=\fontdimen2\font plus
\BIBentryALTinterwordstretchfactor\fontdimen3\font minus \fontdimen4\font\relax}
\providecommand{\BIBforeignlanguage}[2]{{%
\expandafter\ifx\csname l@#1\endcsname\relax
\typeout{** WARNING: IEEEtran.bst: No hyphenation pattern has been}%
\typeout{** loaded for the language `#1'. Using the pattern for}%
\typeout{** the default language instead.}%
\else
\language=\csname l@#1\endcsname
\fi
#2}}
\providecommand{\BIBdecl}{\relax}
\BIBdecl

\bibitem{bouvier2014kabini}
D.~Bouvier, B.~Cohen, W.~Fry, S.~Godey, and M.~Mantor, ``Kabini: An amd accelerated processing unit system on a chip,'' \emph{IEEE Micro}, 2014.

\bibitem{elster2022nvidia}
A.~C. Elster and T.~A. Haugdahl, ``{NVIDIA Hopper GPU and Grace CPU highlights},'' \emph{Computing in Science \& Engineering}, 2022.

\bibitem{abdulah2018exageostat}
S.~Abdulah, H.~Ltaief, Y.~Sun, M.~G. Genton, and D.~E. Keyes, ``{ExaGeoStat}: A high performance unified software for geostatistics on manycore systems,'' \emph{IEEE Transactions on Parallel and Distributed Systems}, vol.~29, no.~12, pp. 2771--2784, 2018.

\bibitem{higham_mary_2022}
N.~J. Higham and T.~Mary, ``Mixed precision algorithms in numerical linear algebra,'' \emph{Acta Numerica}, 2022.

\bibitem{li2019evaluating}
A.~Li, S.~L. Song, J.~Chen, J.~Li, X.~Liu, N.~R. Tallent, and K.~J. Barker, ``Evaluating modern {GPU} interconnect: {PCIe}, {NVLink}, {NV-SLI}, {NVSwitch} and {GPU}direct,'' \emph{IEEE Transactions on Parallel and Distributed Systems}, 2019.

\bibitem{appelhans2017leveraging}
D.~Appelhans and B.~Walkup, ``Leveraging {NVLink} and asynchronous data transfer to scale beyond the memory capacity of {GPU}s,'' in \emph{Workshop on Latest Advances in Scalable Algorithms for Large-Scale Systems}, 2017.

\bibitem{chien2019performance}
S.~Chien, I.~Peng, and S.~Markidis, ``Performance evaluation of advanced features in cuda unified memory,'' in \emph{IEEE/ACM Workshop on Memory Centric High Performance Computing}, 2019.

\bibitem{chu2020nv}
C.-H. Chu, P.~Kousha, A.~A. Awan, K.~S. Khorassani, H.~Subramoni, and D.~K. Panda, ``{NV}-group: link-efficient reduction for distributed deep learning on modern dense {GPU} systems,'' in \emph{ACM International Conference on Supercomputing}, 2020.

\bibitem{fusco2024understanding}
L.~Fusco, M.~Khalilov, M.~Chrapek, G.~Chukkapalli, T.~Schulthess, and T.~Hoefler, ``Understanding data movement in tightly coupled heterogeneous systems: A case study with the {Grace Hopper} superchip,'' \emph{arXiv preprint arXiv:2408.11556}, 2024.

\bibitem{rennich2016accelerating}
S.~C. Rennich, D.~Stosic, and T.~A. Davis, ``Accelerating sparse {Cholesky} factorization on {GPU}s,'' \emph{Parallel Computing}, 2016.

\bibitem{ghysels2022high}
P.~Ghysels and R.~Synk, ``High performance sparse multifrontal solvers on modern {GPU}s,'' \emph{Parallel Computing}, 2022.

\bibitem{foley2017ultra}
D.~Foley and J.~Danskin, ``Ultra-performance {Pascal} {GPU} and {NVLink} interconnect,'' \emph{IEEE Micro}, 2017.

\bibitem{Wilkinson-mp}
\BIBentryALTinterwordspacing
L.~Fox, H.~D. Huskey, and J.~H. Wilkinson, ``{Notes on the Solution of Algebraic Linear Simultaneous Equations},'' \emph{The Quarterly Journal of Mechanics and Applied Mathematics}, vol.~1, no.~1, pp. 149--173, 01 1948. [Online]. Available: \url{https://doi.org/10.1093/qjmam/1.1.149}
\BIBentrySTDinterwordspacing

\bibitem{Moler-mp}
\BIBentryALTinterwordspacing
C.~B. Moler, ``Iterative refinement in floating point,'' \emph{J. ACM}, vol.~14, no.~2, p. 316–321, apr 1967. [Online]. Available: \url{https://doi.org/10.1145/321386.321394}
\BIBentrySTDinterwordspacing

\bibitem{Dongarra-mp}
\BIBentryALTinterwordspacing
J.~J. Dongarra, C.~B. Moler, and J.~H. Wilkinson, ``Improving the accuracy of computed eigenvalues and eigenvectors,'' \emph{SIAM Journal on Numerical Analysis}, vol.~20, no.~1, pp. 23--45, 1983. [Online]. Available: \url{https://doi.org/10.1137/0720002}
\BIBentrySTDinterwordspacing

\bibitem{haidar2018harnessing}
A.~Haidar, S.~Tomov, J.~Dongarra, and N.~J. Higham, ``Harnessing {GPU} tensor cores for fast {FP16} arithmetic to speed up mixed-precision iterative refinement solvers,'' in \emph{International Conference for High Performance Computing, Networking, Storage and Analysis}, 2018.

\bibitem{amestoy_2021}
P.~Amestoy, A.~Buttari, N.~Higham, J.-Y. L'Excellent, T.~Mary, and B.~Vieuble, ``Five-precision {GMRES}-based iterative refinement,'' University of Manchester, Tech. Rep., 2021.

\bibitem{abdelfattah2021survey}
A.~Abdelfattah, H.~Anzt, E.~G. Boman, E.~Carson, T.~Cojean, J.~Dongarra, A.~Fox, M.~Gates, N.~J. Higham, X.~S. Li, J.~Loe, P.~Luszczek, S.~Pranesh, S.~Rajamanickam, T.~Ribizel, B.~F. Smith, K.~Swirydowicz, S.~Thomas, S.~Tomov, T.~Y. M, and U.~M. Yang, ``A survey of numerical linear algebra methods utilizing mixed-precision arithmetic,'' \emph{The International Journal of High Performance Computing Applications}, vol.~35, pp. 344--369, 2021.

\bibitem{abdelfattah2019fast}
A.~Abdelfattah, S.~Tomov, and J.~Dongarra, ``Fast batched matrix multiplication for small sizes using half-precision arithmetic on {GPU}s,'' in \emph{IEEE International Parallel and Distributed Processing Symposium}, 2019.

\bibitem{abdelfattah2020investigating}
------, ``Investigating the benefit of {FP16}-enabled mixed-precision solvers for symmetric positive definite matrices using {GPU}s,'' in \emph{International Conference on Computational Science}, 2020.

\bibitem{higham2021exploiting}
N.~J. Higham and S.~Pranesh, ``Exploiting lower precision arithmetic in solving symmetric positive definite linear systems and least squares problems,'' \emph{SIAM Journal on Scientific Computing}, vol.~43, pp. A258--A277, 2021.

\bibitem{hogg2010fast}
J.~D. Hogg and J.~A. Scott, ``A fast and robust mixed-precision solver for the solution of sparse symmetric linear systems,'' \emph{ACM Transactions on Mathematical Software}, vol.~37, 2010.

\bibitem{abdulah2024boosting}
S.~Abdulah, A.~H. Baker, G.~Bosilca, Q.~Cao, S.~Castruccio, M.~G. Genton, D.~E. Keyes, Z.~Khalid, H.~Ltaief, Y.~Song \emph{et~al.}, ``Boosting earth system model outputs and saving petabytes in their storage using exascale climate emulators,'' \emph{arXiv preprint arXiv:2408.04440}, 2024.

\bibitem{ltaief2024toward}
H.~Ltaief, R.~Alomairy, Q.~Cao, J.~Ren, L.~Slim, T.~Kurth, B.~Dorschner, S.~Bougouffa, R.~Abdelkhalak, and D.~E. Keyes, ``Toward capturing genetic epistasis from multivariate genome-wide association studies using mixed-precision kernel ridge regression,'' \emph{arXiv preprint arXiv:2409.01712}, 2024.

\bibitem{furrer2006covariance}
R.~Furrer, M.~G. Genton, and D.~Nychka, ``Covariance tapering for interpolation of large spatial datasets,'' \emph{Journal of Computational and Graphical Statistics}, vol.~15, no.~3, pp. 502--523, 2006.

\bibitem{sang2012full}
H.~Sang and J.~Z. Huang, ``A full scale approximation of covariance functions for large spatial data sets,'' \emph{Journal of the Royal Statistical Society Series B: Statistical Methodology}, vol.~74, no.~1, pp. 111--132, 2012.

\bibitem{stein2004approximating}
M.~L. Stein, Z.~Chi, and L.~J. Welty, ``Approximating likelihoods for large spatial data sets,'' \emph{Journal of the Royal Statistical Society Series B: Statistical Methodology}, vol.~66, no.~2, pp. 275--296, 2004.

\bibitem{fuentes2007approximate}
M.~Fuentes, ``Approximate likelihood for large irregularly spaced spatial data,'' \emph{Journal of the American Statistical Association}, vol. 102, no. 477, pp. 321--331, 2007.

\bibitem{sun2012geostatistics}
Y.~Sun, B.~Li, and M.~G. Genton, ``Geostatistics for large datasets,'' in \emph{Advances and challenges in space-time modelling of natural events}.\hskip 1em plus 0.5em minus 0.4em\relax Springer, 2012, pp. 55--77.

\bibitem{abdulah2019geostatistical}
S.~Abdulah, H.~Ltaief, Y.~Sun, M.~G. Genton, and D.~E. Keyes, ``Geostatistical modeling and prediction using mixed precision tile {Cholesky} factorization,'' in \emph{IEEE International Conference on High Performance Computing, Data, and Analytics}, 2019.

\bibitem{abdulah2021accelerating}
S.~Abdulah, Q.~Cao, Y.~Pei, G.~Bosilca, J.~Dongarra, M.~G. Genton, D.~E. Keyes, H.~Ltaief, and Y.~Sun, ``Accelerating geostatistical modeling and prediction with mixed-precision computations: A high-productivity approach with {PaRSEC},'' \emph{IEEE Transactions on Parallel and Distributed Systems}, vol.~33, pp. 964--976, 2021.

\bibitem{salvana2022parallel}
M.~L.~O. Salva{\~n}a, S.~Abdulah, H.~Ltaief, Y.~Sun, M.~G. Genton, and D.~E. Keyes, ``Parallel space-time likelihood optimization for air pollution prediction on large-scale systems,'' in \emph{Proceedings of the Platform for Advanced Scientific Computing Conference}, 2022, pp. 1--11.

\bibitem{cao2022framework}
Q.~Cao, R.~Alomairy, Y.~Pei, G.~Bosilca, H.~Ltaief, D.~Keyes, and J.~Dongarra, ``A framework to exploit data sparsity in tile low-rank {Cholesky} factorization,'' in \emph{IEEE International Parallel and Distributed Processing Symposium}, 2022.

\bibitem{cao2023reducing}
Q.~Cao, S.~Abdulah, H.~Ltaief, M.~G. Genton, D.~Keyes, and G.~Bosilca, ``Reducing data motion and energy consumption of geospatial modeling applications using automated precision conversion,'' in \emph{IEEE International Conference on Cluster Computing}, 2023.

\bibitem{onesided-static}
\BIBentryALTinterwordspacing
J.~Kurzak, H.~Ltaief, J.~Dongarra, and R.~M. Badia, ``Scheduling dense linear algebra operations on multicore processors,'' \emph{Concurrency and Computation: Practice and Experience}, vol.~22, no.~1, pp. 15--44, 2010. [Online]. Available: \url{https://onlinelibrary.wiley.com/doi/abs/10.1002/cpe.1467}
\BIBentrySTDinterwordspacing

\bibitem{twosided-static}
\BIBentryALTinterwordspacing
H.~Ltaief, J.~Kurzak, J.~Dongarra, and R.~M. Badia, ``Scheduling two-sided transformations using tile algorithms on multicore architectures,'' \emph{Sci. Program.}, vol.~18, no.~1, p. 35–50, Jan. 2010. [Online]. Available: \url{https://doi.org/10.1155/2010/574728}
\BIBentrySTDinterwordspacing

\bibitem{potrfgpu}
H.~Ltaief, S.~Tomov, R.~Nath, P.~Du, and J.~Dongarra, ``A scalable high performant cholesky factorization for multicore with gpu accelerators,'' in \emph{High Performance Computing for Computational Science -- VECPAR 2010}, J.~M. L.~M. Palma, M.~Dayd{\'e}, O.~Marques, and J.~C. Lopes, Eds.\hskip 1em plus 0.5em minus 0.4em\relax Berlin, Heidelberg: Springer Berlin Heidelberg, 2011, pp. 93--101.

\bibitem{gneiting2010matern}
T.~Gneiting, W.~Kleiber, and M.~Schlather, ``Mat{\'e}rn cross-covariance functions for multivariate random fields,'' \emph{Journal of the American Statistical Association}, vol. 105, pp. 1167--1177, 2010.

\bibitem{pan2024gpu}
Q.~Pan, S.~Abdulah, M.~G. Genton, D.~E. Keyes, H.~Ltaief, and Y.~Sun, ``{GPU}-accelerated vecchia approximations of gaussian processes for geospatial data using batched matrix computations,'' in \emph{ISC High Performance 2024 Research Paper Proceedings (39th International Conference)}.\hskip 1em plus 0.5em minus 0.4em\relax Prometeus GmbH, 2024, pp. 1--12.

\bibitem{agullo2009numerical}
E.~Agullo, J.~Demmel, J.~Dongarra, B.~Hadri, J.~Kurzak, J.~Langou, H.~Ltaief, P.~Luszczek, and S.~Tomov, ``{Numerical linear algebra on emerging architectures: The PLASMA and MAGMA projects},'' in \emph{Journal of Physics: Conference Series}, vol. 180, no.~1.\hskip 1em plus 0.5em minus 0.4em\relax IOP Publishing, 2009, p. 012037.

\end{thebibliography}

\end{document}